# MindGPT: Advancing Human-AI Interaction with Non-Invasive fNIRS-Based Imagined Speech Decoding

Suyi Zhang[,a], Ekram Alam[a], Jack Baber[a], Francesca Bianca[,a], Edward Turner[,a], Maysam Chamanzar[a,b,c,d], Hamid Dehghani[a,e]

[a] *MindPortal Ltd, London, EC2A 3PG, United Kingdom*
[b] *Department of Electrical and Computer Engineering, Carnegie Mellon University, Pittsburgh, PA 15213, USA*
[c] *Department of Biomedical Engineering, Carnegie Mellon University, Pittsburgh, PA 15213, USA*
[d] *Neuroscience Institute, Carnegie Mellon University, Pittsburgh, PA 15213, USA*
[e] *School of Computer Science, University of Birmingham, Birmingham B15 2TT, United Kingdom*

| **Dr Suyi Zhang** | **Dr Francesca Bianco** | **Edward Turner** | **Ekram Alam** | **Jack Baber** |
|---|---|---|---|---|
| Senior Research Engineer | Senior Research Scientist | Research Engineer | CEO & Concept Architect | CTO & Technical Architect |

| **Prof Maysam Chamanzar** | **Prof Hamid Deghani** |
|---|---|
| Scientific Advisor | Scientific Advisor |

**Abstract:** In the coming decade, artificial intelligence systems are set to revolutionise every industry and facet of human life. Building communication systems that enable seamless and symbiotic communication between humans and AI agents is increasingly important. This research advances the field of human-AI interaction by developing an innovative approach to decode imagined speech using non-invasive high-density functional near-infrared spectroscopy (fNIRS). Notably, this study introduces MindGPT, the first thought-to-LLM (large language model) system in the world.

This study focuses on enhancing human-AI communication by utilising fNIRS data to develop a proprietary AI model called MindGPT capable of decoding imagined speech. Hemodynamic responses representing neural activity were collected from four participants instructed to imagine three different sentences. An Extra Trees Classifier (XTC) model was employed to decode neural patterns and differentiate imagined speech from rest conditions, achieving an average accuracy of ~66% across participants, with the best average accuracy at 71%. To further decode neural signals associated with specific imagined sentences, a convolutional neural network (CNN) and a ridge regression model were used as decoders. The CNN model demonstrated an advantage with minimally preprocessed optical density data, outperforming the ridge regression model in this task.

Our results showed significant decoding accuracy for imagined speech, with the ridge regression model achieving a best accuracy of 57% for one participant (chance level: 33%, p-value < 0.001) and the CNN model achieving 47% (chance level: 33%, p-value < 0.001). This study expands our understanding of semantic representation within the brain and supports the crucial role of the dorsolateral prefrontal cortex (DLPFC) in imagined speech processing.

To showcase the practical implementation of our findings, we developed a proof-of-concept near real-time AI communication system using fNIRS technology and a Flask application (i.e., MindGPT), enabling early-stage thought-based communication between participants and the OpenAI GPT-4 API. The implications of this direct communication channel extend across various fields, offering exciting opportunities for human-AI interaction.

By advancing our knowledge of imagined speech processing and demonstrating the potential of fNIRS-based AI communication systems, this study highlights the transformative possibilities of this technology, potentially shaping the future of neurotechnology and AI communication through the monitoring of brain function using fNIRS. Future work will focus on improving decoder accuracy by incorporating a wider range of semantic meanings and employing more advanced machine learning techniques.

# 1.Introduction

Human-AI interaction and human-computer interaction could be radically enhanced if humans had the direct capability to transmit their thoughts to artificial intelligence. The concept of humans imagining language in their mind without any physical involvement is sometimes referred to as imagined speech. Therefore decoding imagined speech using an AI model trained on brain data of participants imagining speech offers the potential for a natural and intuitive means of communication. An imagined speech system could make life better for people who have difficulty communicating and also revolutionise consumer technology and interaction with future AI assistants.
To date there has been some research into extracting neural representations of imagined speech.

Non-invasive neuroimaging techniques traditionally used to investigate the neural underpinnings of imagined speech include electroencephalography (EEG) (with a review by Lopez-Bernal et al., 2022), magnetoencephalography (MEG) (e.g., Dash et al., 2020; Orpella et al., 2022), and functional magnetic resonance imaging (fMRI) (Huth et al., 2016; Tang et al., 2023). These modalities have been successful in revealing temporal as well as spatial patterns associated with language processing but are less precise when it comes to measuring brain activity at high spatial resolution (EEG) or practical application in real-world environments (MEG and fMRI). Presently, invasive methods including electrocorticography (ECoG) have exhibited precision rates for speech comprehension at levels upwards of 60% (e.g., Moses et al., 2019). However, the usage of ECoG is constrained as it demands surgery, and this feature increases doubts regarding its acceptability by users and safety.

Functional near-infrared spectroscopy (fNIRS) has been considered as a promising method of neuroimaging for BCI applications involving imagined speech decoding. fNIRS is a non-invasive technique that measures blood oxygenation changes in the brain as an indirect measure of neural activity, through monitoring the associated vascular responses (slow signal), similar to that of fMRI (Naseer & Hong, 2015). In contrast to EEG, fNIRS provides better spatial resolution and is less sensitive to motion artefacts, allowing it to be applied in more naturalistic settings (Quaresima & Ferrari, 2019). fNIRS, being portable, harmless and cost-effective, as well as not requiring conductive gels or any elaborate set-up procedures, provides a number of advantages (Naseer and Hong, 2015). High-density fNIRS systems, and high-density diffuse optical tomography (HD-DOT) as an extension, can provide better coverage and have been shown to report similar neuroimaging capabilities to fMRI (e.g., Cao et al., 2018; Eggebrecht et al., 2012) with better temporal resolution given the higher sampling rate. This advancement is significant, as it aligns with and extends the growing body of work exploring the application of fMRI in decoding thoughts and semantic information (Tang et al., 2023).

Despite the potential advantages of fNIRS, its application in imagined speech decoding remains relatively underexplored compared to other neuroimaging modalities. Previous studies have demonstrated the feasibility of using fNIRS to decode perceived speech (e.g.,

Liu and Ayaz, 2018) and speech decoding associated with reading (e.g., Hofmann et al., 2014). However, research specifically focused on decoding imagined speech using fNIRS is limited. Cao et al. (2018) and Rybář et al. (2021) have used fNIRS to investigate the decoding of semantic categories of imagined concepts, highlighting the potential of fNIRS for capturing semantic information. The decoding of single words using fNIRS has also been previously explored, e.g., by Naseer and Hong (2014) who developed an online binary decision ("yes", "no") decoding system. These studies provide compelling evidence of the potential of using fNIRS for semantic decoding; however, they focus on single-word decoding, this way limiting the decoding of larger imagined speech extracts. To our knowledge, only Tang et al. (2023) have reported successful identification of 5 1-min stories from brain data; however, the authors used fMRI neuroimaging method, which is renowned for its high spatial resolution. Nevertheless, no study has systematically investigated the use of fNIRS for decoding imagined sentences.

The current study aims to address this gap in the literature by exploring the feasibility of using fNIRS to decode imagined speech compared to rest condition and in decoding one of three imagined sentences in a classification task. We propose a novel fNIRS-based system that combines advanced signal processing techniques and machine learning algorithms to classify imagined sentences with different semantic content from a predefined set, based on the analysis of brain activity patterns. The primary objective of this study is to achieve above-chance decoding accuracy in detecting imagined speech, demonstrating the potential of fNIRS as a viable modality for imagined sentence decoding. Different levels of fNIRS data preprocessing were also implemented and tested to investigate which data preparation method would lead to best model performance. This decision was driven by the fact that previous studies (e.g., Eastmond et al., 2022) suggested that deep learning can be used to reduce lengthy preprocessing. In addition, we aim to identify brain regions associated with imagined speech, and those specific to the different imagined sentences, to extend our knowledge of semantic representation in the brain (Brumberg et al., 2016; Eggebrecht et al., 2014; Petersen et al., 1988). Finally, we present a proof-of-concept near real-time fNIRS-based system that enables direct human-AI interaction through imagined speech. More specifically, the sentences decoded by our developed models are then utilised as prompts for a language model, i.e., ChatGPT (Achiam et al., 2023), to facilitate telepathy-like and direct human-AI communication.

In summary, this paper presents a novel approach for decoding imagined speech using high-density fNIRS and crucially shows the first demonstration of direct imagined sentences interaction with an LLM. We demonstrate successful decoding of imagined sentences as the first step towards implementing a practical and non-invasive solution for imagined speech decoding through the following innovations: (i) Demonstrating the feasibility of fNIRS recording using off-the-shelf commercially available headgears for collecting high signal-to-noise ratio (SNR) signals during imagined speech, compared to rest brain function; (ii) Implementing decoding algorithms that can decipher and classify imagined sentences from a limited dictionary with relatively high accuracy; (iii) Extending our knowledge of semantic representation in the brain; and (iv) Establishing an early-stage thought-based communication channel as a platform for new possibilities in human-AI interaction and synergy. This BCI paradigm can be further optimised and refined for intriguing applications involving human-AI interaction, beyond imagined speech decoding.

# 2.Material and Methods

## 2.1 Participants

This study included a total of 4 participants (male = 3, female = 1), ranging in age from 25 to 28 years, with a mean age of 26.5 years. Participants were required to be fluent in English, free from any neurological disorders, and not currently undergoing any form of psychological treatment. Participants were also asked to refrain from drinking coffee and make use of any other substance that may alter their state before each data collection session. Participants wearing glasses were asked to use contact lenses for data collection for ease when setting up the fNIRS neuroimaging cap. All participants provided informed consent and were compensated for their time. All participants completed data collection in less than 9 sessions.

## 2.2 Experimental Setup

In this study, we employed a Continuous-Wave (CW) high-density 48x48 fNIRS system that provides full-head coverage. The commercially available fNIRS system (NIRx Inc.), consisted of 48 sources and 47 detectors (the extra detector is used for the short-distance channels). The system consists of a total of 388 channels (194 source wavelength at 760 nm and 194 source wavelength at 850 nm); sampling rate: 5.9 Hz; channel distances ranging from ~21 mm to ~42 mm), and 8 short-distance channels (channel distances: < 10 mm) providing high-density full-head coverage (see *Figure 1*), to monitor changes in oxygenated blood levels in the brain as a proxy for neural activity.

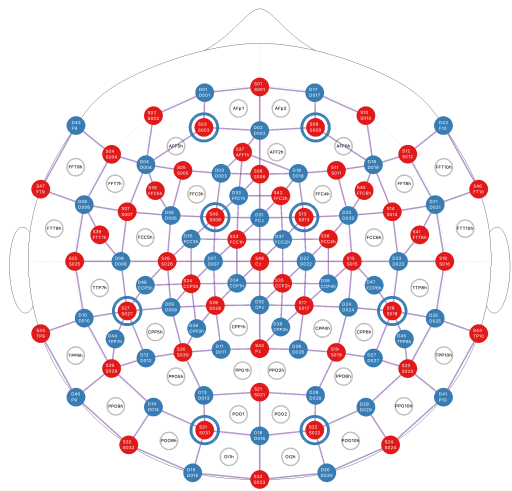

*Figure 1. Whole head high-density fNIRS montage, including 48 sources, 47 detectors and 8 short-distance channels.*

Participants were asked to sit in front of a computer screen, with their hands resting on the table, and the fNIRS fibre bundles arranged in a ponytail attached to the main fNIRS box. An example setup is illustrated in *Figure 2*. Participants memorised three sentences before the onset of the experiment. At the start of the experiment, participants were reminded using a beeping metronome tone to imagine the sentences at a set pace, while their brain was recorded with fNIRS. Using a metronome ticking at 100 beats per minute as a reference pace for the imagined sentences ensured consistent timing and speed with the intention for participants to imagine the sentences at 100 words per minute.

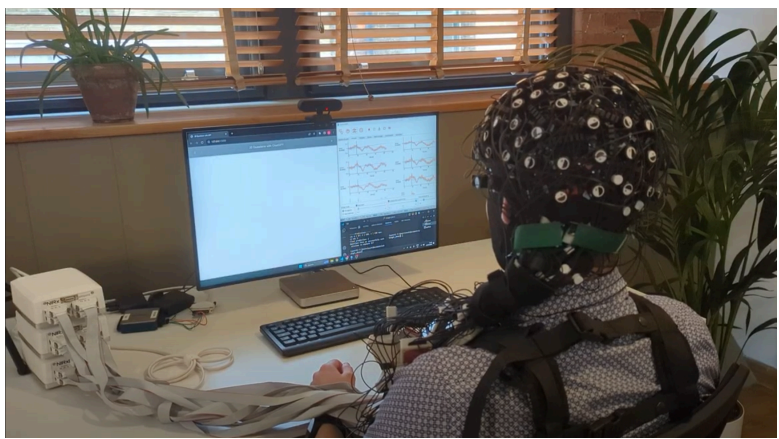

*Figure 2. Example of a participant imagining the sentences during a fNIRs brain scanning session*

To avoid an effect of sentence duration on decoding performance, the sentences used in the experiment were carefully designed to have similar lengths of approximately 25 seconds when imagined. Moreover, the selected sentences (reported in *Table 1* below) were crafted to have distinct semantic content (BERT score calculation between sentences was used as a measure of semantic similarity (Zhang et al., 2019)). To obtain a robust dataset, each sentence was imagined repeatedly by participants in a randomised order. The total number of imagined speech trials collected from participants were: 423 for participant 1, 378 for participant 2, 423 for participant 3, and 419 for participant 4. These repetitions were collected across 11 to 16 separate experimental sessions for each participant. All imagined speech trials included an equivalent number of the three imagined sentences (e.g., the 423 imagined speech trials collected from participant 1 included 141 trials for each of the three imagined sentences). Rest condition trials were also collected (n = 25 for participant 1, n = 126 for participant 2, n = 141 for participant 3, and n = 140 for participant 4) with the same duration (25s) as imagined sentence trials. Please note that the number of rest condition trials collected from participant 1 is significantly lower compared to the other participants. This is due to the fact that rest condition trials were implemented in our paradigm when participant 1 had already started data collection, while all other participants started data collection after this point and thus completed a higher number of rest condition trials. These were presented to participants in a randomised order and were used in the brain activation analyses. During rest condition trials, the participant was instructed to clear their mind and refrain from doing or thinking anything in particular.

*Table 1. Imagined sentences with distinct semantic content selected for the 3-class classification*

| Prompt | Imagined Sentence |
|---|---|
| Call | I have a great time speaking to Mom on the phone, and I know that she does too. As we live so far away from each other, texting just does not do it justice. I think giving her a call is the only way to fully catch up. |
| Restaurant | We have not had a date night in forever, and I would love to try that new Italian place that opened downtown. There is always a big queue outside the restaurant, let me just make a reservation so we do not miss out. |
| Venus | If I were on Venus, I'd be in a world of extremes. The pressure here feels like being a kilometre underwater, crushing me from all sides. The air is a corrosive nightmare, capable of dissolving metal. |

## 2.3 fNIRS Data Preprocessing

In this work, we tested different levels of preprocessing of our raw fNIRS data to identify which data preparation method would lead to best model performance. With this choice, we aimed to illustrate the balance between a lengthy preprocessing of fNIRS data to extract relevant information to decode imagined speech and the complexity of the models used. More specifically, while CNNs are more complex models compared to a ridge regression model, the former is renown for being able to extract features from minimally preprocessed data (e.g., Kumar et al., 2023), while a higher level of preprocessing may be needed for the latter. We therefore reasoned that raw or minimally preprocessed (optical density) data might result in higher performance when selecting the CNN decoder, while fully preprocessed data might be more suitable for the ridge regression, simpler model. Initially, the brain data collected during sentence imagining was streamed through the NIRx acquisition software, Aurora fNIRs (NIRx Medical Technologies LLC), and saved in the XDF format. The saved data was then loaded and subjected to a series of preprocessing techniques. Specifically, we trained our models using (a) raw data, i.e., intensity only, (b) optical density data, i.e., minimally preprocessed data which underwent conversion of raw signals to optical density data [-log(Intensity_task/Intensity_rest)], and (c) fully preprocessed data, i.e., data underwent conversion of raw signals to optical density, detrending, short channel regression correction, motion artefact correction, conversion to haemoglobin concentration using a partial pathlength factor (ppf) of 6, and bandpass filtering between 0.01 and 0.7 Hz (Huppert et al., 2009). Finally, regardless of the level of preprocessing, data was trimmed to a fixed shape of 145 time points by 388 channels (optical density and fully preprocessed data: 194 oxy-haemoglobin and 194 deoxy-haemoglobin channels) and saved as HF5 files, with the filenames corresponding to the respective sentence names.

## 2.4 Imagined Speech Detection Decoder

In this study, we applied an Extra Trees Classifier (XTC) model to fully preprocessed brain signals to decode imagined speech. This model was implemented using the ExtraTreesClassifier from the scikit-learn library (Geurts, et al., 2006) with the following parameters: bootstrap = False; criterion = "entropy"; max_features = 0.2; min_samples_leaf = 6; min_samples_split = 7; n_estimators = 100. The XTC is an ensemble learning method based on the random forest algorithm, which fits multiple decision trees on various sub-samples of the dataset and uses averaging to improve predictive accuracy and control over-fitting. A stratified k-fold cross-validation with 3 folds was employed and tests were conducted over 5 different seeds to ensure the model's robustness and versatility capability. Stratified sampling maintained the class distribution across folds. Decoder performances were assessed using decoding accuracy as the metric of success, which is defined as the accuracy of classification of the predicted test set over many trials. To assess the significance of the classification results, a p-value was calculated using the cumulative distribution function (CDF) of the binomial distribution. The p-values from each fold were combined using Chi-squared distribution to obtain a single p-value representing the overall statistical significance of the results. Class-wise accuracy distribution was also tracked to analyse the model's performance across different classes (imagined speech vs rest condition). All models were trained on single-subjects to determine best accuracy per subject, although results from subjects were also averaged to identify overall performance of our models in successfully completing imagined speech detection. Participant 1 was excluded from this test, as not enough rest condition trials were collected to allow for this classification.

## 2.5 Imagined Sentences Decoder

In this study, we employed advanced machine learning techniques to decode brain signals and classify the sentences that participants were imagining into one of the 3 predefined classes. We developed and implemented two decoding models to achieve this goal. The first model was based on a one-dimensional Convolutional Neural Network (1D-CNN) architecture to analyse time-series fNIRS data. In addition, we used the widely used ridge regression model for imagined sentence classification as the baseline. Ridge regression models have been used in previous language-related brain decoding studies (Huth et al., 2016, Tang et al., 2023), which can be used as baseline models to compare with our innovative CNN model. Both models can be used in the final near real-time demonstration (see section 2.4). All models were trained on single-subjects to determine best accuracy per participant, although results from participants were also averaged to identify overall performance of our models in successfully completing imagined sentence classification across participants. As a slightly different number of trials was collected from participants, the total number of trials included per participant considered are following: (1) Participant 1: 423 imagined speech trials, (2) Participant 2: 378 imagined speech trials, (3) Participant 3: 423 imagined speech trials; (4) Participant 4: 419 imagined speech trials. All imagined speech trials include an equivalent number of the three imagined sentences (e.g., Participant 1 has a total of 423 imagined speech trials, which include 141 trials for each of the three imagined sentences). We do not expect this variation among participant trials to greatly affect the performance of our decoding algorithms, as overall a similar number of trials were collected for all participants.

Decoder performances were assessed using decoding accuracy as the metric of success, which is defined as the accuracy of classification of predicted test set over many trials. The imagined sentence and rest condition labels were selected as the ground truth. A 3-fold cross validation was run 5 times with different random seeds. We reported average and best fold accuracy by seed and participant, and we combine those values to determine a single value for participants' average and best accuracy across different seeds. For each fold, a p-value is calculated using the cumulative distribution function (CDF) of the binomial distribution. The combined p-value of all 3 folds across seeds is then calculated using a Fisher's test (see Supplementary Information for detailed results).

### 2.5.1 1D-CNN Model

A 1D-CNN with multiple layers was developed for each participant (subject-specific) for the analysis of time series fNIRS data. Similarly, 1D-CNNs have been previously used in BCI applications to decode fNIRs brain signals to remove the requirement for complex brain image processing (e.g., Kumar et al., 2023), as they are quite effective at extracting features from noisy signals. However, to the best of our knowledge, direct application of 1D-CNN to decode imagined speech has not been demonstrated before. The 1D-CNN represented a good balance between model complexity and effectiveness at extracting relevant features for imagined speech decoding, thus they were preferable to the more complex 2D- or 3D-CNNs. Regularisation techniques, including dropout and fully connected layers, were incorporated into our model. Xavier initialization (Glorot & Bengio, 2010) was employed for weight initialization, and following softmax activation (for two-class classification for imagined speech vs rest condition) or sigmoid activation (for three-class classification for the three imagined sentences), the final layer's output was determined.

The recorded fNIRS data was partitioned into training, validation, and test subsets. The scikit-learn's Robust scaler was fitted with training data and applied to validation and test data separately. The model was trained and subsequently evaluated on the validation set before final performance assessment on the test set. The Cross Entropy Loss was utilised as the loss function, while the Adam optimizer was employed for optimization steps. The learning rate was optimised using the ReduceLROnPlateau scheduler. Furthermore, L1 and L2 regularisation, early stopping, and cross-validation were implemented, along with a search for L1 and L2 coefficients. Model performance was evaluated using the accuracy of the predicted test set, where imagined sentence and rest condition labels were selected as ground truth. This same procedure was followed for all data, regardless of the preprocessing steps taken to prepare the data, and for all participants.

### 2.5.2 Ridge regression model

A cross-validated ridge regression model was also implemented for each participant (subject-specific) as the baseline model to convert fNIRS brain signals to sentence embeddings. Scikit-learn's RidgeCV was used with the alpha range set to $10e^{-3}$ - $10e^{3}$ (following Tang et al., 2023) where the best alpha value that determines regularisation strength is determined with cross validation. The sentence embeddings were created using SentenceTransformer, where each sentence is encoded into 768-element 1D vectors. The base model 'all-mpnet-base-v2' was used. The predicted embeddings from ridge regression is then fed into a logistic regression classifier to output the predicted identity of the sentence.

For pre-processing, the fNIRS data was scaled with scikit-learn's Robust scaler. The first 10 samples of each trial were averaged to obtain the baseline, which was then subtracted from the rest of the signal for removing the baseline. Delta Hb/HbO value over 16 (µMol for haemo data) was clamped in order to remove outliers. The scaler was fitted with training data and applied to validation and test data separately. This same procedure was followed for all data, regardless of the preprocessing steps taken to prepare the data, and for all participants.

## 2.6 Imagined speech related brain activations

In addition to applying decoding models directly to the imagined speech data, analyses involving haemodynamic response modelling and GLM-based statistical testing were conducted to identify imagined speech-related activations in the brain. To identify semantic representation in the brain, brain activations during imagined speech and rest condition were compared (fully preprocessed fNIRS data were used; see section 2.3). This was repeated for all participants. During our experiments, a slightly different number of trials was collected for each participant: (1) Participant 1: 423 imagined speech vs 25 rest condition trials, (2) Participant 2: 378 imagined speech vs 126 rest condition trials, (3) Participant 3: 423 imagined speech vs 141 rest condition trials; (4) Participant 4: 419 imagined speech vs 140 rest condition trials. All imagined speech trials included an equivalent number of the three imagined sentences (e.g., the 423 imagined speech trials collected from participant 1 included 141 trials for each of the three imagined sentences). GLM works best when multiple trials can be averaged, therefore, we expect that the few rest condition trials collected for participant 1 might have a negative impact on this analysis. However, we do not expect this variation among the other participant trials to affect our analyses greatly, as a high number of trials was consistently collected for participants 2, 3, and 4.

The fNIRS data was fully preprocessed using the steps described in section 2.3. A first-level design matrix is constructed to model the hemodynamic response associated with neural activity, using the python package mne-nirs (version 0.6.0). Processed haemo data was obtained by converting raw fNIRS signals to optical density signals, and then into haemo data using the Beer Lambert law with ppf=6. Channels (distance > 10mm) are used in the general linear model (GLM, Friston et al., 1994), with a cosine function to model and correct for low-frequency drift in the signal, and a high-pass filter of 0.005Hz to remove slow signal variations not contributed to neuronal activities. The adopted hemodynamic response function (HRF) model is based on the Statistical Parametric Mapping (SPM) approach, a standard model for estimating the brain's vascular response to neural activity. Lastly, different stimulus durations were considered to assess the temporal progression of brain activations throughout the 25s interval (durations of 5s, 10s, 15s, 20s, and 25s) (please note that temporal progression of brain activations was not generated for participant 1 due to computer memory insufficiency to process data saved in the used file format - data was saved in a more efficient data format for all other participants). While the full brain activation results are outlined in *Supplementary Figure S1*, we report the most characteristic activation for each participant in our results section. Short channel data (distance < 10mm) are included as nuisance regressors in the design matrix. Conditions 'Imagined speech' and 'rest condition' are specified and the GML parameters are estimated. The contrast 'Imagined speech > rest condition' is then estimated from GLM theta values, and z-scores are calculated. Surface plots are generated with the estimated z-scores.

To verify the accuracy of our method to localise brain areas recruited during the cognitive tasks and compare this accuracy in localisation across participants, we also conducted finger tapping experiments at the start and end of each session and compared brain activation in response to right vs left finger tapping and their localisation within the cortex (please note that brain activation visualisation for the finger tapping task was not developed for participant 1, as the finger tapping task was added to our paradigm after this participant completed data collection). Right vs left finger tapping tasks have been previously shown to lead to a quite strong and clear differential activation in the contralateral motor cortex, with respect to the hand completing the finger tapping (e.g., Batula et al., 2017). Therefore, identifying the correct localisation of brain activation during the finger tapping experiment would confer us with increased confidence with respect to precise cap placement and brain region coverage, as well as good quality of data collected. See *Supplementary Figure S2* for a brief explanation of the finger tapping paradigm used, as well as the brain activations.

## 2.7 Illustrative MindGPT Application

To demonstrate a practical application of our fNIRS-based system, we consider a scenario where our BCI enables early-stage human-AI thought-based communication, which we refer to as MindGPT. For demonstration purposes, we developed an innovative Flask-based application aimed to show a proof-of-concept use case that automates sending decoded human thoughts to ChatGPT. The core of our MindGPT system involved integration of brain activity data, captured during sentence imagery tasks, with the capabilities of the OpenAI GPT4 API (OpenAI, 2023) for a direct mind to OpenAI communication. The initial phase of our workflow was the acquisition of brain data via the NIRx acquisition software (Aurora fNIRs) which was subsequently stored in the SNIRF format. This raw data underwent a series of preprocessing steps described in section 2.3.

Leveraging the preprocessed fNIRS data, we evaluated various models to facilitate effective interaction with OpenAI GPT4. The 1d-CNN model served as the foundation for classification tasks. This setup enabled dynamic retrieval of corresponding texts through a dictionary lookup mechanism, with the decoded sentence being stored, subsequently fetched by a Flask server, and presented within a web interface. Interactions with the OpenAI GPT4 API were driven by these queries, with the system capable of receiving and displaying responses to users in near real-time (*Figure 3*). We define our application as 'near real-time' as a slight delay is present between new data collection from a user and the decoded output, resulting from the need to start and stop the data recording system between decoding attempts. This limitation is driven by the NIRx acquisition software (Aurora fNIRs) used to collect fNIRS data, but we aim to improve the seamlessness in the future to allow for a traditional real-time application. GPT4 was instructed to provide useful suggestions based on the user's imagined input, in an attempt to create a coherent dialogue.

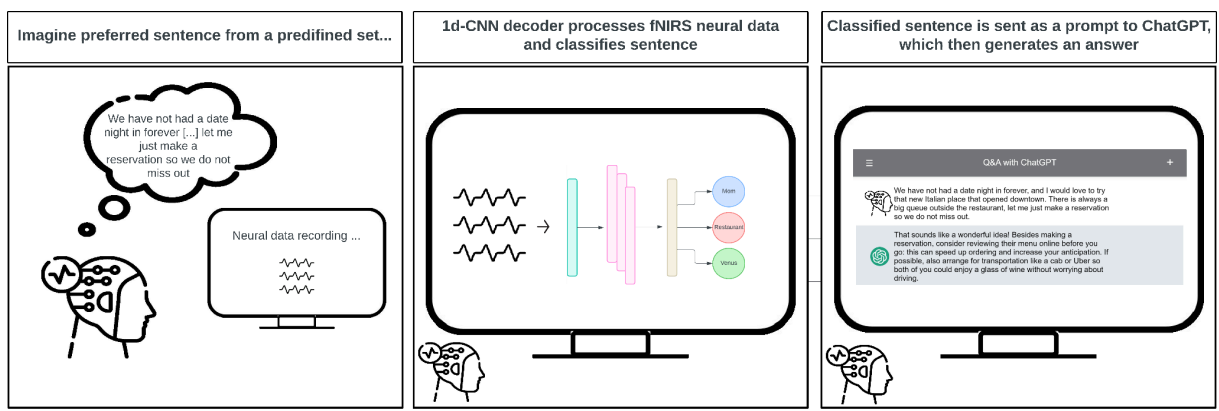

*Figure 3. Illustration of MindGPT application. A user first imagines their preferred sentences from a predefined set of 3 sentences, while their brain data is being recorded using fNIRS. In this example, the participant is imagining the sentence related to going to the 'Restaurant'. Then, the participant's brain data is provided as input into a 1d-CNN decoder which processes fNIRS data and classifies the data into one of the 3 sentence options ('Call', 'Restaurant', 'Venus'). The text associated with the classified sentence is then retrieved from a look up table and sent as a prompt to ChatGPT in the Flask application. ChatGPT then generates an answer based on the user's imagined input, attempting this way to create a coherent dialogue. Please note this process happens in near real-time (see text for details).*

To demonstrate the potential and impact of MindGPT for implementing an effective communication link between humans and machines, we showcase the application of MindGPT for human dialogue with OpenAI GPT-4 (OpenAI, 2023) (*Figure 4*). For each imagined sentence, 10 related questions to the specific topic were generated using ChatGPT itself prior to the MindGPT experiment, each of which was also converted to embeddings using SentenceTransformer. After decoding the first sentence and sending it to GPT4, the next round of dialogue was introduced by comparing the ridge regression-generated sentence embeddings. The closest matching ChatGPT generated question to the imagined sentence, measured by cosine similarity of sentence embeddings, is then picked from a lookup table and sent to GPT4 again to continue the conversation. This is similar to a zero-shot learning approach (Palatucci et al., 2009), where new test cases are unseen by the trained model.

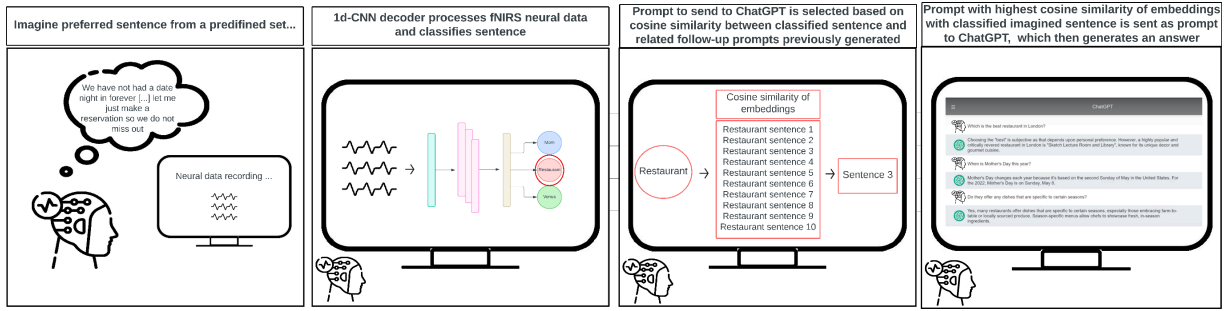

*Figure 4. Illustration of continuous MindGPT application. A user first imagines their preferred sentences from a predefined set of 3 sentences, while their brain data is being recorded using fNIRS. In this example, the participant is imagining the sentence related to going to the 'Restaurant'. Then, the participant's brain data is provided as input into a 1d-CNN decoder which processes fNIRS data and classifies the data into one of the 3 sentence options ('Call', 'Restaurant', 'Venus'). The closest matching ChatGPT-generated question from a list of 10 questions to the classified imagined sentence (i.e., Restaurant in this case), measured by cosine similarity of sentence embeddings, is then picked from a lookup table and sent to GPT4 to create a coherent dialogue. These steps are repeated multiple times to allow for a continued conversation. Please note this process happens in near real-time (see text for details).*

# 3. Results

## 3.1 Imagined Speech Decoding vs Rest Condition

In this section we report the performance of our XTC model in detecting imagined speech from the rest condition when classifying neurovascular signals associated with the two conditions (see *Figure* 5). A total of 162 imagined speech and 162 rest condition trials were included for participants 2 and 3, while participant 4 counted 123 imagined speech and 123 rest condition trials due to time constraints. As mentioned previously, participant 1 was excluded from this analysis since not enough rest condition trials were collected from participant 1 to train and test a decoder (n = 25 rest condition trials only). Overall, our XTC model achieved an average accuracy of ~66% (p-value < 0.001) when considering averaged accuracies across folds in the 3 subjects included in this test. Our best participant (participant 1) reported a best average accuracy across the 3 folds of ~71% (p-value < 0.001). See *Supplementary Table S1* for detailed results.

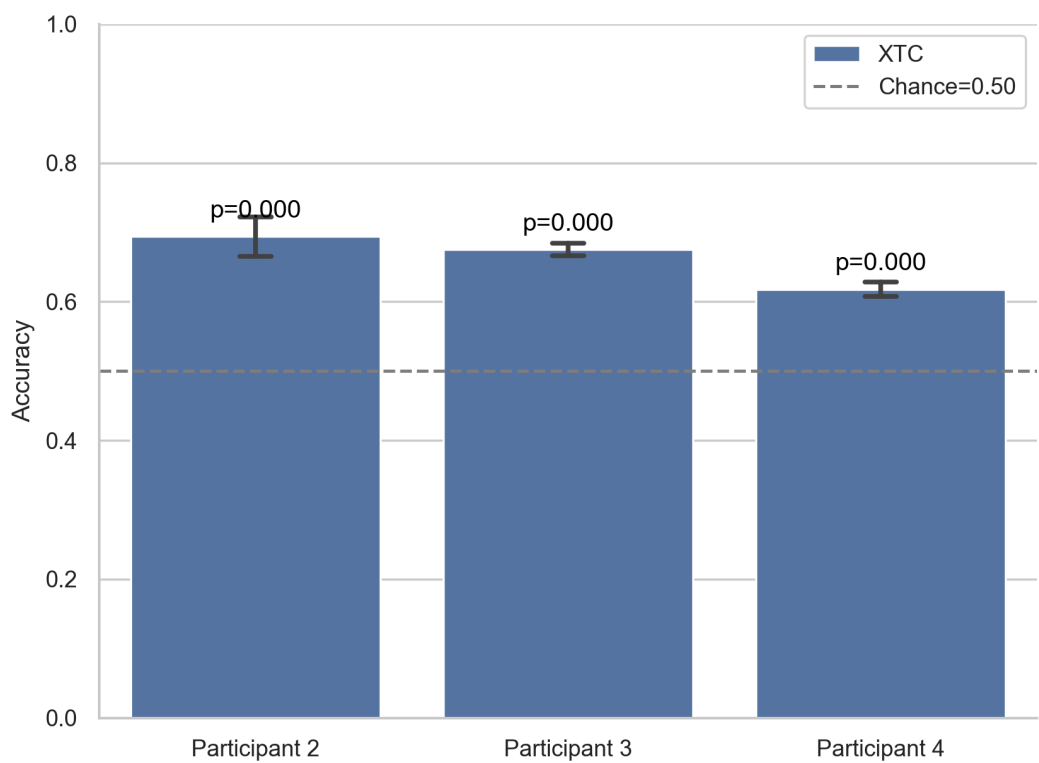

*Figure 5. Summary of XTC model performance (accuracy %) in classifying imagined speech vs the* rest *condition when using fully preprocessed fNIRS data. Data, standard deviation and p-values are reported for participants 2, 3, and 4. Comparison to chance (50%) is also reported.*

## 3.2 Imagined Sentence Decoding

In this section, we report and compare the performance of our two models in decoding brain data into one of the 3 predefined imagined sentence classes across participants. We also compare the accuracy of our 1d-CNN and ridge regression-based language models, when using different preprocessing steps for our raw fNIRS data, in order to identify which level of preprocessing leads to the best performance. *Table 2* shows that our models were able to decode all participants' fNIRS neural signals associated with different imagined sentences and identify which of the three predefined sentences the participants were imagining to a significant level (chance level: 33%) (see *Supplementary Tables S2-S7* for detailed results). Best accuracy was achieved in participant 2 with the language model-based decoder, when using fully preprocessed data (~57.4% accuracy, chance: 33%, p < 0.001). When comparing results more in detail across participants, the 1D-CNN was seen to outperform the ridge regression-based language model-based decoder in 2 out of 4 participants (participants 1 and 4), whereas the opposite was valid for participants 2 and 3. Interestingly, highest accuracy using the 1D-CNN was achieved when data underwent minimal preprocessing, i.e., conversion of raw signals to optical density data. In contrast, highest accuracy using the Ridge Regression was obtained when fully preprocessed data was used. No preprocessing achieved worse performance in both models, compared to both minimally and fully preprocessed data. See *Table 2* and *Supplementary Tables S2-S7 for detailed results*.

Table 2. Summary of 1D-CNN and language model decoder performance in decoding imagined brain data into one of the three predefined sentences. Averaged accuracies across different folds and seeds were reported for each participant. These are presented by (a) model used - 1D-CNN or Ridge Regression-based language model; and (b) data type - Raw: no preprocessing, OD: optical density (or minimal preprocessed data), Preproc: preprocessed data (or fully preprocessed data with pipeline described). Standard deviation and statistical testing are reported in Supplementary Tables S2-S7. The highest accuracy level achieved using each of the two models (1D-CNN and Ridge Regression-based language model) is shown in bold-face numbers for each participant. Finally, the winning model was reported for each participant, indicating the model that achieved highest accuracy for that specific participant.

|  | 1D-CNN | | | Ridge Regression-based language model | | | Winning Model |
| --- | --- | --- | --- | --- | --- | --- | --- |
| Participant | Raw | OD | Preproc | Raw | OD | Preproc | |
| Participant 1 | 0.355 | **0.418** | 0.398 | 0.355 | **0.404** | 0.397 | **1D-CNN** |
| Participant 2 | 0.315 | 0.454 | **0.467** | 0.405 | 0.504 | **0.574** | **Ridge Regression** |
| Participant 3 | 0.303 | 0.396 | **0.400** | 0.359 | 0.379 | **0.433** | **Ridge Regression** |
| Participant 4 | 0.308 | **0.457** | 0.419 | 0.389 | 0.377 | **0.393** | **1D-CNN** |

### 3.2.1 1D-CNN Model

A total number of 423, 378, 423, and 419 imagined speech trials were collected from participants 1, 2, 3, and 4, respectively. All imagined speech trials included an equivalent number of the three imagined sentences (e.g., the 423 imagined speech trials collected from participant 1 included 141 trials for each of the three imagined sentences). The model was trained on 70% of all data (11-16 sessions based on the participant considered), while the test dataset included 30% of all trials. When using raw data, the 1D-CNN model achieved an average accuracy of 32.0% across the 4 participants included in this study (averaged highest fold accuracy 35.5% in participant 2), therefore performing around chance level (33.3%). The p-value did not achieve significance in any of the participants (see *Supplementary Tables S2*). In contrast, when using minimally preprocessed (i.e., optical density) data, the 1D-CNN model achieved an average accuracy of ~43.1% across the 4 participants included in this study (highest average fold accuracy of ~45.4% in participant 4). The p-value achieved significance across 5 tests (p<0.05) in 3 fold cross validations with different seeds in all participants (see *Supplementary Tables S3*). Finally, when using fully preprocessed data, the 1D-CNN model achieved an average accuracy of 42.1% (highest average fold accuracy of ~46.7% in participant 2). Again, the p-value achieved significance across 5 tests (p<0.05) in 3 fold cross validations with different seeds in all participants (see *Supplementary Tables S4*). See *Supplementary Tables S2-S4* for detailed results, including fold and seed-specific accuracies by participant, as well as standard deviations and statistical testing. Overall, the 1D-CNN model seems to perform best when optical density data are used, thus highlighting the advantages of CNNs for extracting features from noisy and minimally processed data.

### 3.2.2 Ridge regression model

The same number of trials were used for training and testing as in the CNN model. When using raw data, the ridge regression model achieved an average accuracy of 37.7% (averaged highest fold accuracy 40.5%), where p-value achieved statistical significance (p<0.05) in 3 out of 4 participants (3 fold cross validations tested 5 times with different seeds) (see *Supplementary Tables S5*). In contrast, when using minimally preprocessed (i.e., optical density) data, the ridge regression model achieved an average accuracy of 41.6% (averaged highest fold accuracy 44.7%), where p-value achieved high significance (p<0.05) for all participants (see *Supplementary Tables S6*). Finally, when using fully preprocessed data (see section 2.3), the ridge regression model achieved an average accuracy of 44.9% (averaged highest fold accuracy 48.7%), where p-value achieved significance (p<0.05) for all participants (see *Supplementary Tables S7*). Fully preprocessed data achieved the highest accuracy compared to minimally preprocessed (optical density) or raw data. See *Supplementary Tables S5-S7* for detailed results.

## 3.3 Brain areas underlying imagined speech

*Figure 6A* illustrates the surface cortex plots of HbO activations of the contrast between imagined speech and rest condition conditions for all four participants included in this study. While different stimulus durations were considered to assess the temporal progression of brain activation throughout the 25s during which participants imagined each sentence (durations of 5s, 10s, 15s, 20s, and 25s), we here report the most characteristic activation for each participant (see *Supplementary Figure S1* for details on the temporal progression of brain activations*)*. This analysis identified a recurrent brain region recruited across 3 out of 4 participants during imagined speech, i.e., the dorsolateral prefrontal cortex (DLPFC). Interestingly, one of our participants (participant 4) showed a decrease in activation in the dorsolateral prefrontal cortex during imagined speech, which is an unexpected result. We make some speculations in the discussion with respect to this result. Additional regions differentially recruited between participants included the lateral temporal cortex, visual related regions near MT+ complex, auditory cortex, and early sensorimotor cortex. These activations have similarities to early imagined speech studies with PET and fNIRS (*Figure*

*6B and 6C*, Brumberg et al., 2016; Eggebrecht et al., 2014; Petersen et al., 1988), where auditory cortex, DLPFC, frontal eye field, premotor and motor regions were implicated in speech processing. This suggests that fNIRS is capable of capturing imagined speech processing brain activation.

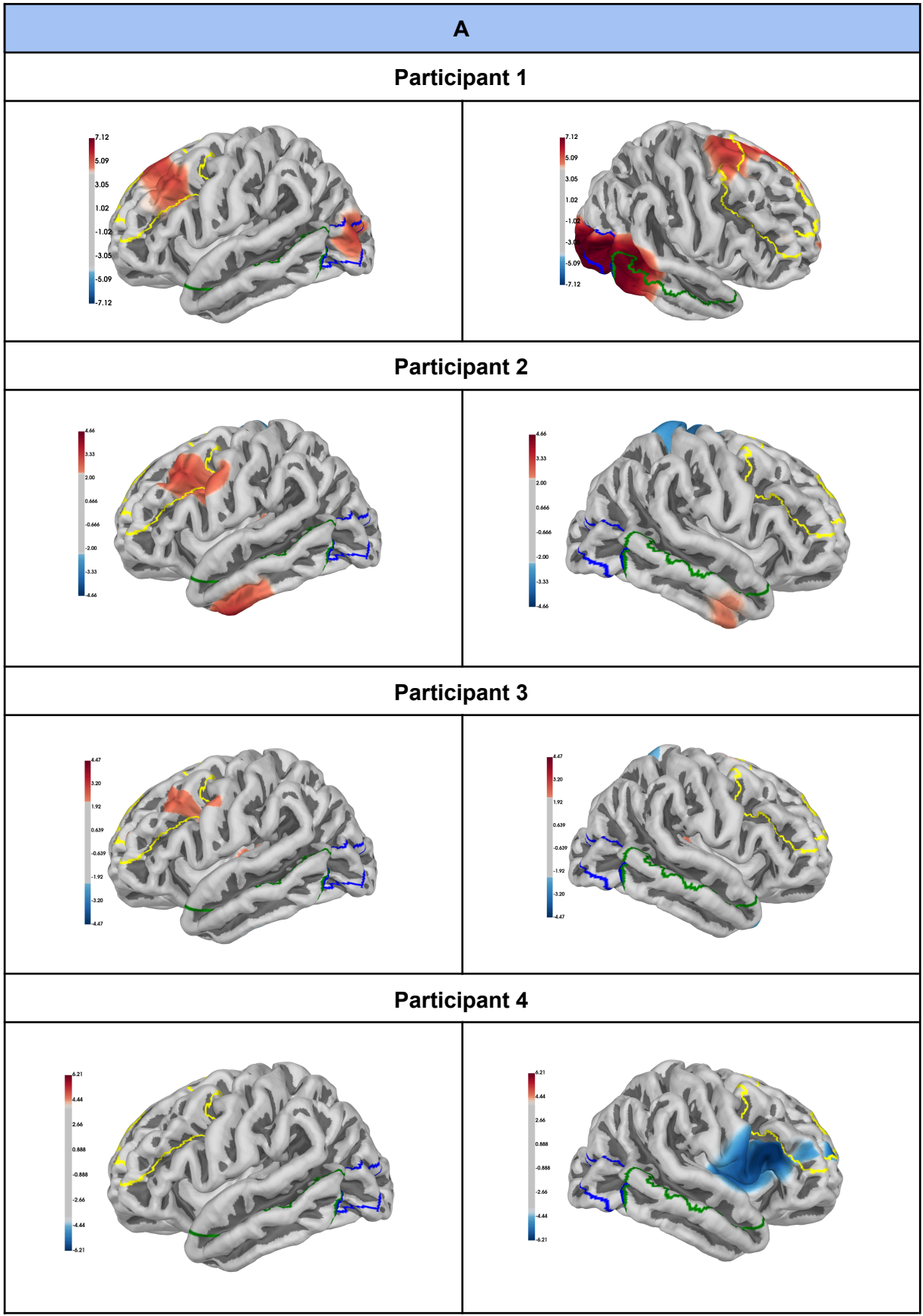

| B |
|---|
| 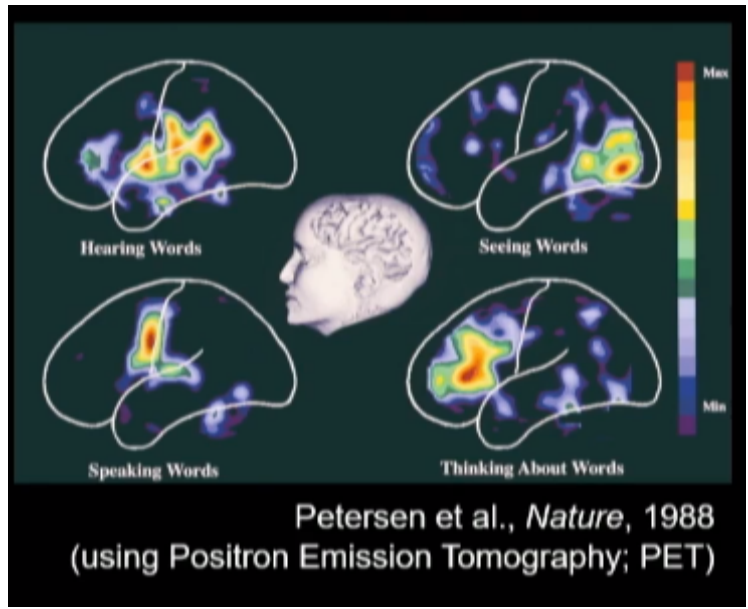 |
| **C** |
| 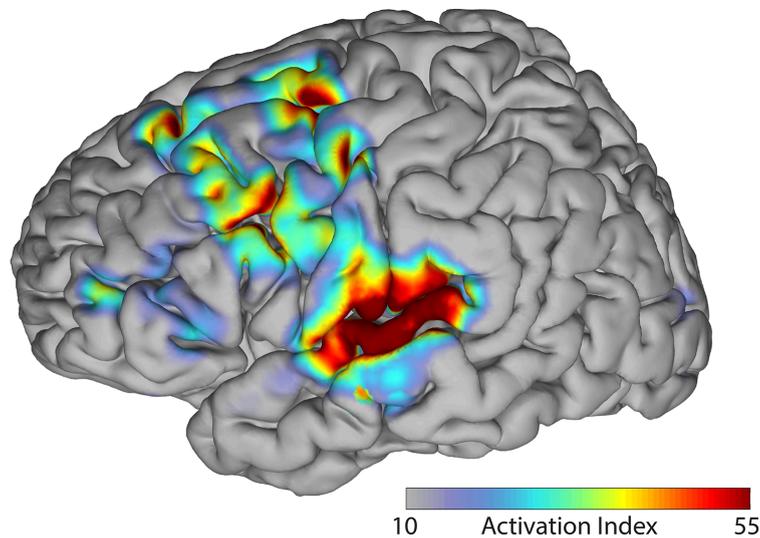 |
| Taken from Brumberg et al., 2016. The spatial topography of statistically significant relationships between brain signals in the overt and covert condition over the entire continuous session. Colour values represent values of activation index (see colour bar). Strong auditory cortex involvement is observed in the superior temporal gyrus, along with lesser activation along a path from the DLPFC, FEF, premotor and motor regions; IFG and vMC sites found in the overt and covert conditions alone are notably absent. https://doi.org/10.1371/journal.pone.0166872.g007 |

*Figure 6. Surface cortex plots showing HbO activations with the contrast Imagined speech > rest condition across all participants. (A) Outlines shown are regions defined in the HCP MMP1 atlas (Glasser et al., 2016), where the green outline is the Lateral Temporal Cortex, the yellow outline is the DorsoLateral Prefrontal Cortex, and the blue outline is the MT+ Complex and Neighboring Visual Areas. The colorbar indicates z-score values. Brain activations reported for all participants during the imagined speech task. (B) Brain activations shown in previous research relating to different forms of speech (Petersen et al., 1988) (Collated by Eggebrecht et al., 2014). (C) Brain activations shown in previous research relating covert vs overt speech processing (Brumberg et al., 2016).*

## 3.4 Illustrative MindGPT Application

To demonstrate a practical application of our fNIRS-based BCI, we consider a scenario where our MindGPT BCI enables early-stage human-AI thought-based communication (see *Figure 3, section 2.7* for a schematic of the mindGPT application). *Figure 7* illustrates an example of MindGTP in action, i.e., a near real-time fNIRS-based BCI which enables telepathy-like human-AI communication. Brain signals are collected and processed in near real-time while the user imagines one of the three predefined sentences (see section 2.7). The data is then decoded by our models (see section 2.5) and classified into one of the three imagined sentences upon which the models were trained. The decoded sentence is then utilised as a prompt for a language model (in this case ChatGPT), enabling telepathy-like communication between the user and ChatGPT. The latency of our MindGPT thought-based communication is 27.62 seconds (CPU: AMD Ryzen 9 5900X 12-Core Processor; GPU: NVIDIA GeForce RTX 3080 Ti; RAM: 64.0 GB). The main bottleneck lies in the extended time required to imagine a sentence (25 seconds), while decoding is pretty fast once the decoder is trained (2.62 seconds - loading fNIRS data file: 0.66 seconds; preprocessing fNIRS data: 0.61 seconds; imagined sentence decoding: 1.34 seconds; trigger sent to ChatGPT: 0.01 seconds). Therefore, the long latency of our near real-time MindGPT is a result of task (imagining sentences) limitations, rather than decoding approach limitations (e.g., long preprocessing needed or long decoding time).

When running instances of MindGPT with our user in near real-time, we observed several errors from our decoders, leading to suboptimal human-AI communication. This is due to the fact that our best decoder (i.e., 1D-CNN) achieved an average accuracy of 42% (see section 3.2.1); therefore, it is bound to commit errors when decoding the sentences the user is imagining. However, the importance of this illustration lies in the instances in which the decoder correctly decodes the sentence the user is imagining, due to the decoder accuracy being significantly above chance (see section 3.2). When imagined sentences are decoded correctly, our MindGPT shows successful early-stage human-AI thought-based communication. This is a significant result as this is the first case of human-AI telepathy-like and direct communication. *Figure 7* below shows successful MindGPT decoding different imagined sentences (pretrained three imagined sentences) from brain data collected from the user at different instances of imagined speech (*Figure 7A, 7B and 7C* show successful decoding of sentences 'Restaurant', 'Call', 'Venus', respectively). See Figure 8B for an example of decoder errors.

| A |
|---|

☰ Q&A with ChatGPT +

We have not had a date night in forever, and I would love to try that new Italian place that opened downtown. There is always a big queue outside the restaurant, let me just make a reservation so we do not miss out.

That sounds like a wonderful idea! Besides making a reservation, consider reviewing their menu online before you go: this can speed up ordering and increase your anticipation. If possible, also arrange for transportation like a cab or Uber so both of you could enjoy a glass of wine without worrying about driving.

| B |
|---|

☰ Q&A with ChatGPT +

I have a great time speaking to Mom on the phone, and I know that she does too. As we live so far away from each other, texting just does not do it justice. I think giving her a call is the only way to fully catch up.

That's lovely! Continuing with regular calls to catch up is indeed meaningful. To make conversations more engaging, you could also consider using video call platforms like Zoom or FaceTime. Not just hearing voices but seeing each other can foster a better connection. You might think about scheduling these calls to have something to look forward to, or even surprise her with random calls when you have good news to share.

| C |
|---|

If I were on Venus, I'd be in a world of extremes. The pressure here feels like being a kilometre underwater, crushing me from all sides. The air is a corrosive nightmare, capable of dissolving metal.

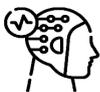 Exploring Venus indeed would be a challenge due to harsh conditions. Extreme pressure, high temperatures, acidic atmospheric composition and sulfuric rainfall makes it inhospitable. If determined to explore, it is highly recommended that all technology, versatility suits or potential robots bear the latest pressure and heat-resistant developments and rigorously-tested shields and filters to guard against extreme elements and corrosive substances. It's critical to ensure all safety considerations before initiating any expedition plan.

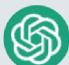

*Figure 7. MindGPT in action. Different sentences (out of a preset of three sentences) are decoded from brain data collected from the user at different instances of imagined speech and inputted as prompts into ChatGPT to allow telepathy-like communication. (A) Sentence "Restaurant" is decoded and inputted into ChatGPT, which starts a conversation about making a reservation; (B) Sentence "Call" is decoded and inputted into ChatGPT, which starts a conversation about regularly calling a mother; and (C) Sentence "Venus" is decoded and inputted into ChatGPT, which starts a conversation about exploring Venus.*

To further illustrate the future capabilities and impact of MindGPT, we implemented a continuous thought-based conversation between our user and ChatGPT, starting from the same three imagined sentences (*Figure 4*). More specifically, we developed 10 different potential follow-up prompts per each imagined sentence topic, e.g., 10 prompts associated with the imagined sentence topic "Restaurant" (see *Table 3* below for an example of generated questions for follow-up of decoded imagined sentence topic "Restaurant"). These follow-up prompts were generated using chatGPT itself prior to the MindGPT experiment. To do this, we prompted ChatGPT with the 3 imagined sentences and asked it to generate 10 extra sentences related to the imagined sentence topics. During near real-time testing, after decoding the imagined sentences from brain signals, one out of the 10 follow-up prompts belonging to the decoded imagined sentence topic was chosen as a prompt to ChatGPT (see *Figure 8*). This was achieved by comparing the ridge regression-generated sentence embedding and selecting the closest matching questions to the generated sentence embedding (cosine similarity) to continue the conversation (see section 2.7 for further details on methodology). This implementation therefore allowed the user to sustain a continued conversation with ChatGPT on the same topic. Similarly to the results reported in the previous paragraph, this human-AI thought-based communication was not optimal (see *Figure 8B* for erroneous continuation of topic decoding), owing to the 42% max accuracy of our models. In addition, hard-coding the input of one of the language model-generated sentences (based on the user's imagined sentence) (see *Table 3* for example follow-up generated questions) also meant that the user had limited control on how the conversation was continued with ChatGPT, apart from setting the 'topic' of conversation by imagining one of the three imagined sentences the decoder was trained on. Regardless of these limitations, this implementation nevertheless exemplifies the first continuous thought-based communication between a human user and AI (ChatGPT).

*Table 3.* Example questions generated for imagined sentence "Restaurant" which are selected as follow-up prompts based on cosine similarity of sentence embedding.

> 1. Are they recognized for any particular beverages such as wines or cocktails?
> 2. What kind of ambiance or setting is the establishment known for?
> 3. Is there a signature dish that stands out on their menu?
> 4. How far in advance are reservations typically required?
> 5. Do they offer any dishes that are specific to certain seasons?
> 6. How does its reputation compare to other establishments in the city?
> 7. Has the restaurant been awarded any culinary accolades?
> 8. Who is the head chef and what is notable about their culinary background?
> 9. How is the overall service quality rated by patrons?
> 10. Is there a recommended dress code for diners?
> 11. Are they recognized for any particular beverages such as wines or cocktails?

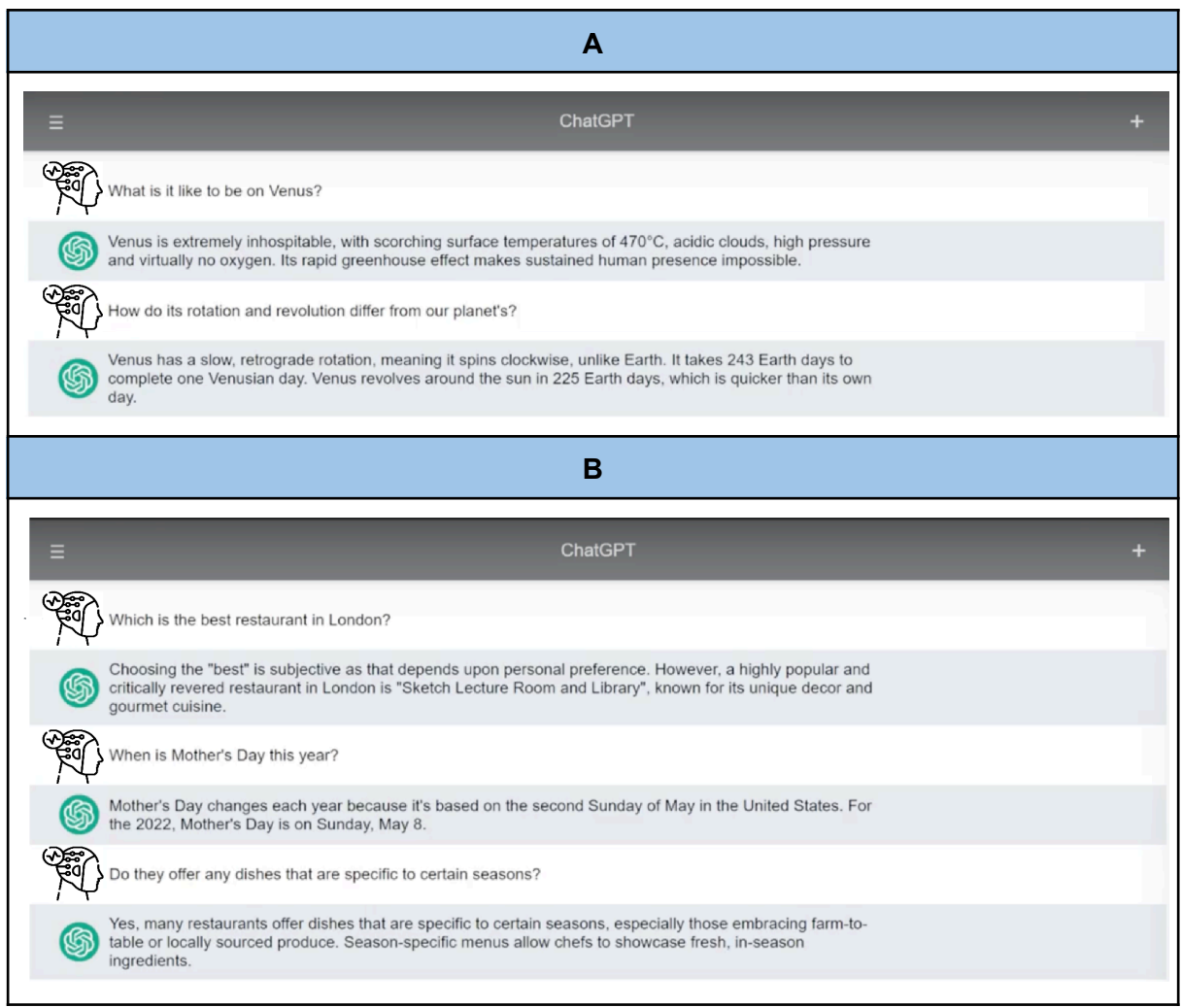

*Figure 8. Continuous MindGPT in action. Different sentences (out of a preset of three sentences) are decoded from brain data collected from the user at different instances of imagined speech and inputted as prompts into ChatGPT to allow telepathy-like communication. To allow for a continuous conversation, after the first decoded imagined sentence, the closest matching question to the generated sentence embedding (cosine similarity) out of ten available questions related to the first decoded imagined sentence is selected as a follow-up prompt to ChatGPT. (A) Example of two correctly decoded imagined sentences in a row ("Venus") and use of a follow-up questions following the method outlined; (B) Example of a correctly decoded imagined sentence ("Restaurant") followed by an erroneously decoded imagined sentence ("Call") and a correctly decoded imagined sentence ("Restaurant").*

Overall, although at its infancy and needing improvement to become a real-world application, MindGPT provides the basis for future human-AI telepathy communication. The successful examples provided in this section were included for the readers to gauge the user experience and impact that our MindGPT would allow once it reached its full capability. The potential applications of MindGPT appear limitless, spanning across various domains and disciplines.

# 4.Discussion

This paper discusses a groundbreaking advance in the field of human-AI interaction and neurotechnology by demonstrating the feasibility of decoding imagined speech and imagined sentences with different semantic meaning using high-density fNIRS. Through the development of an innovative fNIRS pipeline, we achieved above-chance accuracy (average accuracy of ~66%, best accuracy: ~71%, chance level: 50%, p-value <0.001) to determine whether participants (n = 4) engaged in imagined speech or not (i.e., rest condition brain activity). A previous study (Lesaja et al., 2022) showed a ~90% accuracy in detecting overt speech using invasive stereo-EEG. In contrast, our innovative method shows the feasibility of decoding covert (imagined) speech using non-invasive neuroimaging (fNIRS), with a reasonably high level of accuracy. Additionally, we achieved above-chance accuracy (overall best model accuracy: 57% in participant 2, chance level: 33%, p-value < 0.001) in identifying three distinct imagined sentences with different semantic meanings and deployed this capability for human-AI telepathy-like communication in our MindGPT application. This sets our research apart from previous studies that primarily focused on perceived speech (Eggebrecht et al., 2014; Huth et al., 2016; Tang et al., 2023), silent reading (Martin et al., 2014), single-word imagined speech decoding (Naseer and Hong, 2014), or imagined concepts categorisation (e.g., Cao et al., 2018; Rybář et al., 2021). While there has been a previous study on decoding imagined sentences using fMRI (e.g., Tang et al., 2023), our study is the first to successfully decode imagined sentences using a high-density fNIRS system, surpassing the limitations of fMRI in terms of applicability to practical real-world scenarios. We presented the first successful example of a direct human-AI telepathy-like communication, i.e., our MindGPT illustrative demonstration using our imagined speech decoding BCI paradigm.

Our results show that both our models (1D-CNN and ridge regression-based language model) perform above chance to different extents, based on the participant and the level of fNIRS data preprocessing (i.e., raw, optical density, or fully preprocessed data), possibly indicating subject-specific differences to consider when decoding imagined speech from different participants. Our 1D-CNN model outperformed the simpler ridge regression model used in fMRI pipelines (Tang et al., 2023) when extracting semantic features from minimally preprocessed brain data using fNIRS. This suggests that the more complex CNN architecture may better represent the semantic space during imagined speech without significant preprocessing, although further studies are needed to confirm this hypothesis. This is in concordance with previous literature (e.g., Eastmond et al., 2022) suggesting that deep learning can be used to reduce lengthy preprocessing, and again highlights the advantages of using deep learning for extracting features of interest from fNIRS neural data.

Additionally, we identified imagined speech-related brain activations consistent with previous literature (Brumberg et al., 2016; Eggebrecht et al., 2014; Petersen et al., 1988), validating our results and contributing to the understanding of imagined speech representation in the brain. In particular, activities identified in DLPFC highlighted its recently acknowledged importance in language and speech processing (Hertrich et al., 2021). Interestingly, one of our participants (participant 4) showed a decrease in activation in the DLPFC during imagined speech, which is an unexpected result. We could speculate that this result could be attributed to the fact that this participant was a female and that the fNIRS cap was too big for this participant (please note that the same cap was used for all participants), resulting in cap misplacement and noisy data. Indeed, finger tapping brain activation localisation (see *Supplementary Figure S2*) also showed poor results, indicating that data from this participant may require additional processing or feature extraction to be useful.

Our system shows immense promise compared to other neuroimaging tools. EEG has limited spatial resolution and decoding accuracies, while fMRI, despite providing valuable insights, has limited practical applications. In contrast, our system is the first to demonstrate early-stage telepathy-like, direct communication between a human and an AI system, offering a more natural and efficient user experience compared to previous BCI systems that rely on motor activations during imagined speech or employ neuroimaging tools that are difficult to use in real-life scenarios. While our fNIRS-based BCI system needs improvements in decoding accuracy, it shows the rich prospect of thought-based communication between humans and AI.

The field of neurotechnology and its societal impact could be revolutionised by the development of an fNIRS-based system for decoding imagined speech. This technology has the potential to provide individuals with communication disorders an alternative means of expressing their thoughts and intentions, ultimately improving their quality of life and social interactions. Additionally, the integration of imagined speech decoding with AI systems could lead to more natural and intuitive human-machine communication, opening up new possibilities in various domains and potentially propelling human evolution forward. Our work represents a significant step in this direction, demonstrating the feasibility and potential of direct brain-to-AI communication using fNIRS.

In interpreting the results of this study, it is important to be mindful of decoding accuracy, sample size, and ethical considerations. First, a primary limitation with the current system is the accuracy. There are several steps that can be taken to improve the accuracy of the system. These include, collecting more data from more participants, trying a wider range of sentences with a larger variety of semantic meanings and also improving the decoding models. An approach which can potentially achieve these improvements would be the use of a more versatile model that is trained off of a wide range of semantic text rather than just 3 discrete sentences. The versatility of the model would not only allow it to detect a larger corpus of imagined thoughts but might be able to increase the accuracy of decoding by recognising higher order relationships between different semantic thoughts with enough data. To improve the accuracy of our model, we will be conducting future work aimed at collecting additional data and data from additional participants, widening our semantic coverage by moving away from the 3 discrete sentence decoding, and improving our decoding models further.

Second, given this study's reliance on data from only four subjects, we are unable to provide remarks about the generalisability of its findings. Individual variability in neural activity means the current results may not apply broadly, in particular for real-time use cases; future studies should test generalisability concerns directly. To address this limitation, we plan to collect data from additional participants to build a more versatile model and validate our findings from a single-subject.

Finally, when it comes to the development and application of imagined speech decoding systems, there are numerous ethical considerations that need to be taken into account. One of the main concerns is data privacy, as these BCIs have the potential to reveal an individual's private and personal thoughts and intentions. It is crucial to ensure the protection and confidentiality of the collected brain data in order to safeguard the privacy of users. Another potential issue is the possibility for misuse and abuse of BCI technology in the context of imagined speech decryption. While this technology has the potential to greatly benefit human life, it also carries the risk of harm if it is misused. To prevent such abuse and ensure responsible application, it is imperative to establish ethical principles and regulations for the development and utilisation of BCI systems. This will help ensure that the technology is used for its intended purpose and to promote the greater good.

# 5.Conclusion

In conclusion, this study represents a groundbreaking advancement in the field of human-Ai interaction and neurotechnology by demonstrating the feasibility of decoding imagined speech and imagined semantic information using high-density fNIRS. The innovative fNIRS pipeline developed in this research has achieved above-chance accuracy in distinguishing between imagined speech and rest condition (average accuracy of ~66%, best accuracy: ~71%, chance level: 50%, p-value <0.001), as well as different three predetermined imagined sentences (overall best model accuracy: 57% in participant 2, chance level: 33%, p-value < 0.001), enabling its successful implementation in the MindGPT application for human-AI early-stage thought-based communication. In addition, this study supports previously identified brain areas associated with imagined speech and shows the suitability of fNIRS to correctly localise brain areas recruited during such cognitive function.

Despite the notable progress made, it is essential to acknowledge the limitations of this study, such as the need for improved decoding accuracy and the reliance on a small sample size in this study. We will be conducting further research to address these limitations by (a) collecting data from additional subjects to validate and investigate the generalisability of our findings, and (b) improving our decoder accuracy by collecting additional data per participant, covering a wider range of semantic meaning and using more advanced machine learning techniques.

To the best of our knowledge, this study serves as the first demonstration of an fNIRS-based imagined speech decoding paradigm and can be considered as pioneering work, paving the way for future advancements.

# Supplementary Information

**Supplementary Figure S1.** Temporal progression of brain activations throughout the 25s during which participants imagined each sentence (durations of 5s, 10s, 15s, 20s, and 25s). Please note that temporal progression of brain activations was not generated for participant 1 due to computer memory insufficiency to process data saved in the used file format - data was saved in a more efficient data format for all other participants. Brain activation visualisations built using general linear modelling (GLM, Friston et al., 1994).

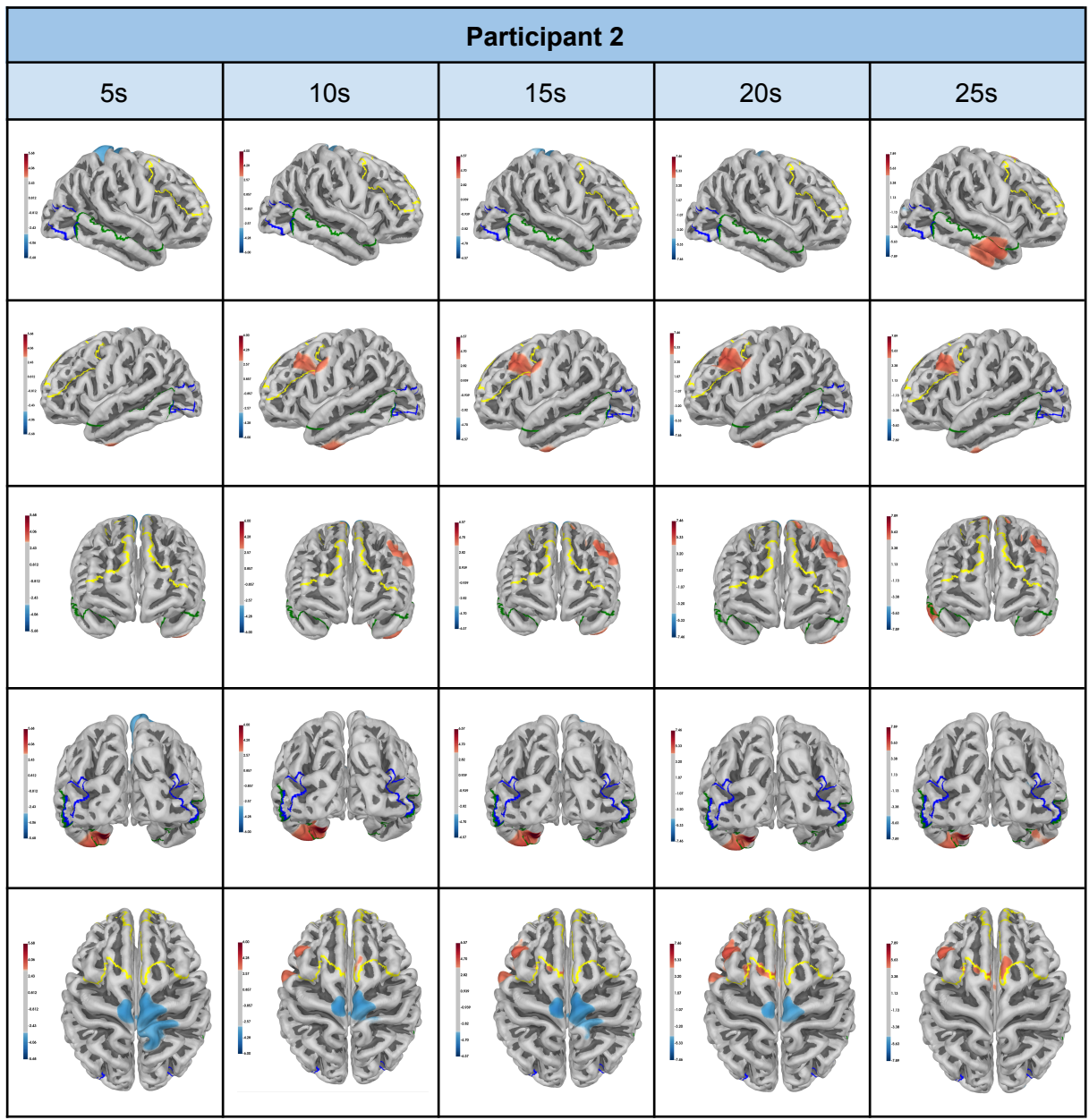

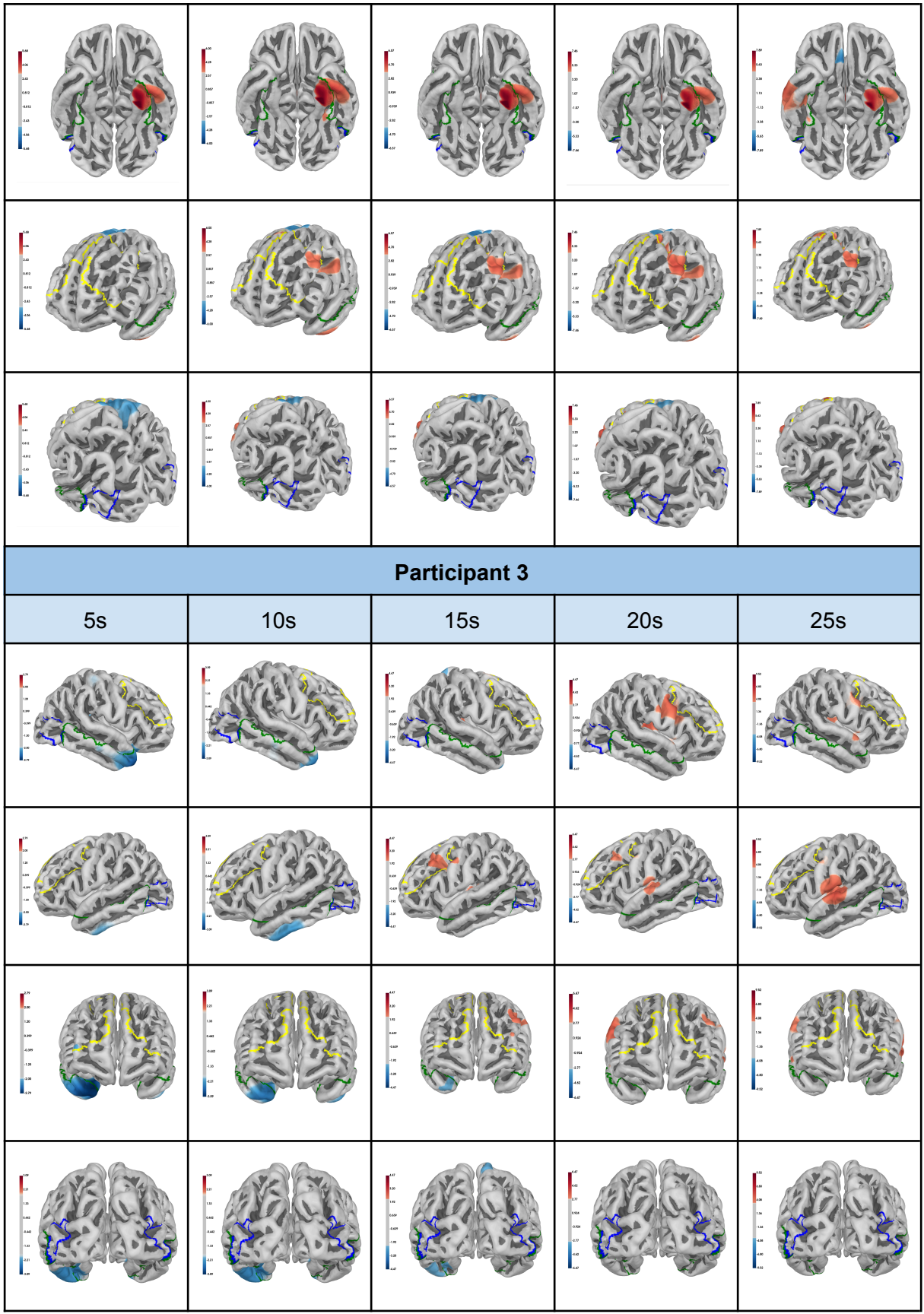

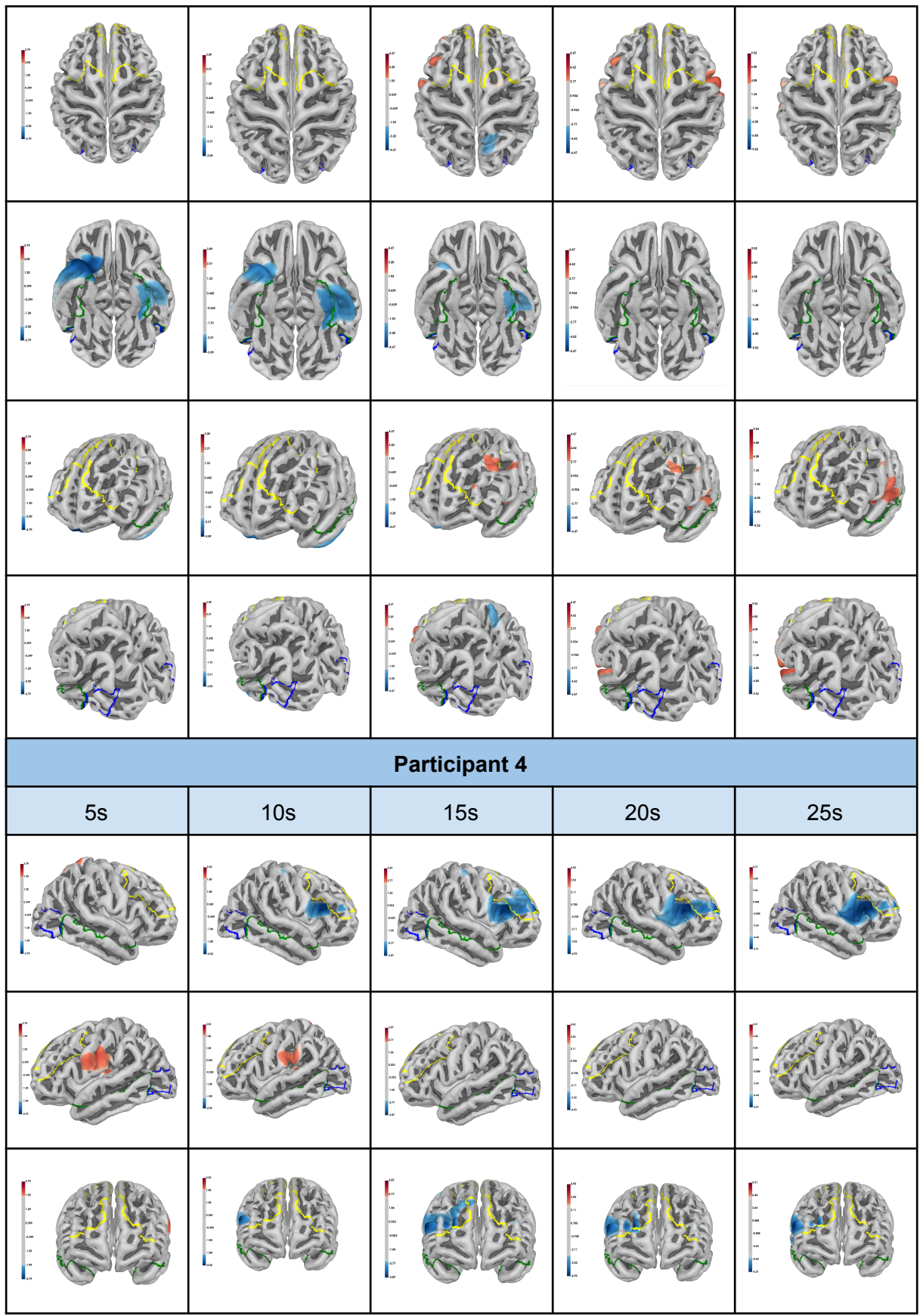

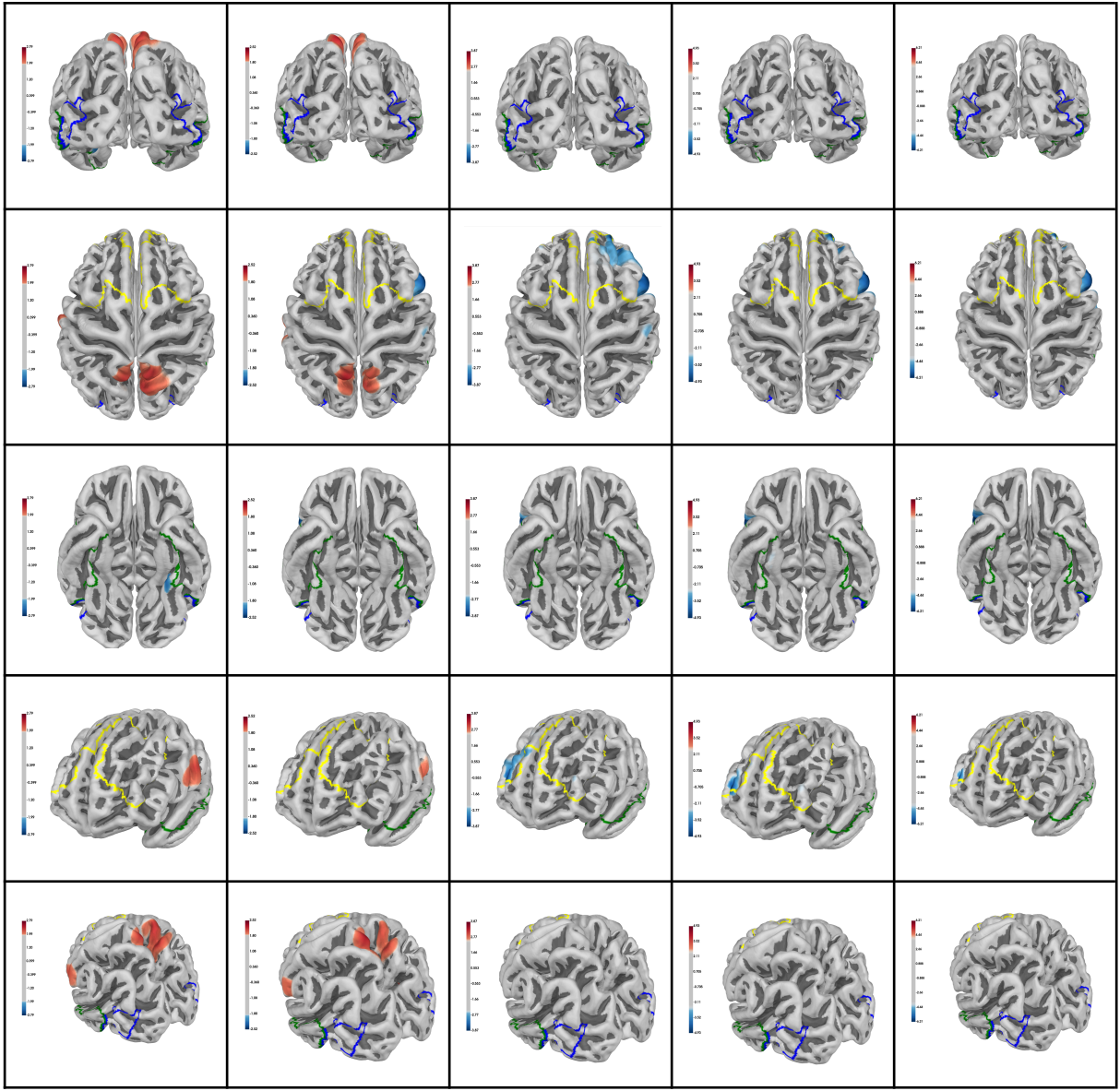

**Supplementary Figure S2.** Description of the finger tapping paradigm conducted at the start and end of each session in this study (A). Participants' brain activations in response to 10s of right or left finger tapping trials and their localisation within the cortex (B-C-D). Please note that brain activation visualisation for the finger tapping task was not developed for participant 1, as the finger tapping task was added to our paradigm after this participant completed data collection. Brain activation visualisations built using general linear modelling (GLM, Friston et al., 1994).

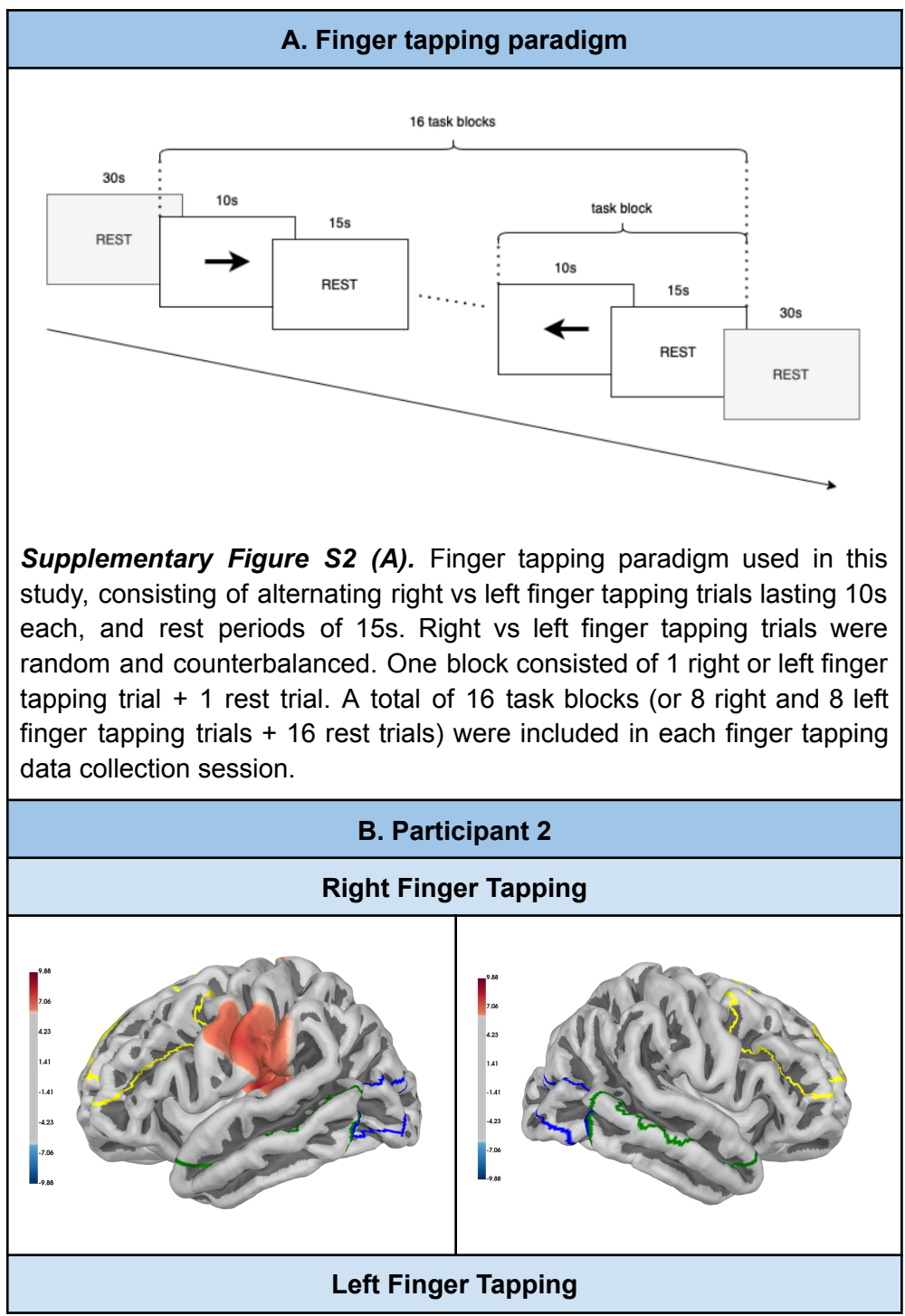

*Supplementary Figure S2 (A).* Finger tapping paradigm used in this study, consisting of alternating right vs left finger tapping trials lasting 10s each, and rest periods of 15s. Right vs left finger tapping trials were random and counterbalanced. One block consisted of 1 right or left finger tapping trial + 1 rest trial. A total of 16 task blocks (or 8 right and 8 left finger tapping trials + 16 rest trials) were included in each finger tapping data collection session.

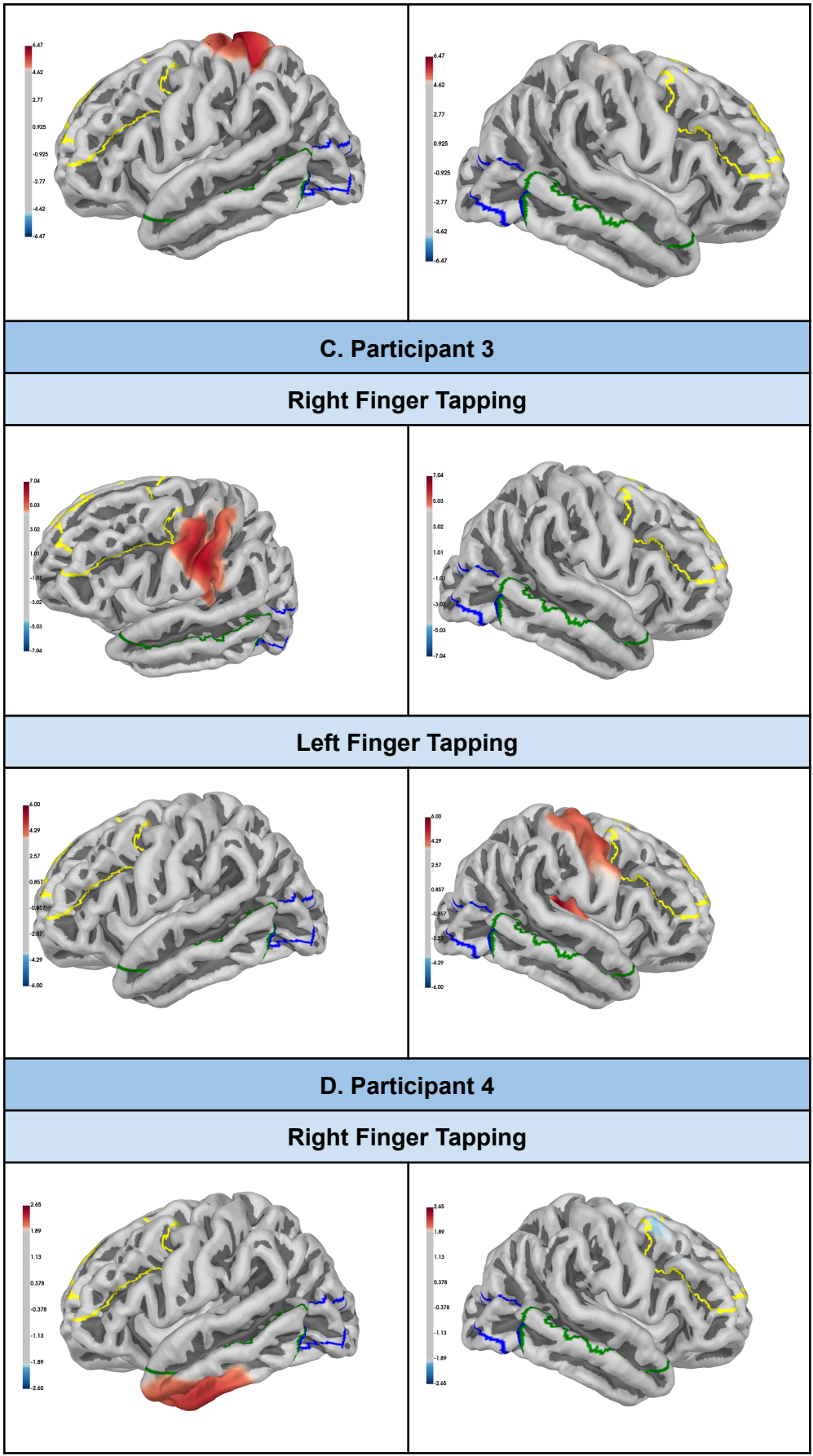

**C. Participant 3**

**Right Finger Tapping**

**Left Finger Tapping**

**D. Participant 4**

**Right Finger Tapping**

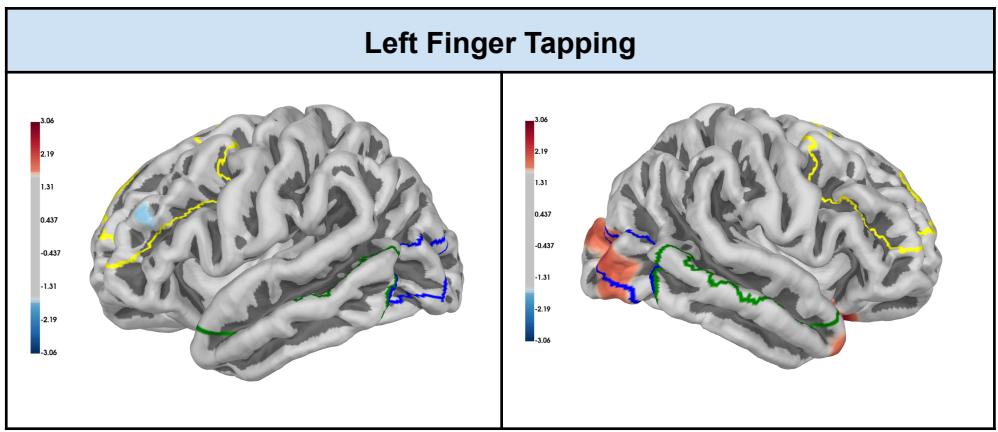

**Supplementary Table S1.** Decoding accuracy performance of the XTC model when inputting fully preprocessed data from participants 2, 3 and 4, assessed over 5 different seeds, and 3 k-folds. Best model accuracy and average model accuracy across the different folds and different seeds per participant are reported; as well as significance (p < 0.05). Averaged values across participants are also outlined.

| Supplementary Table S1. XTC Model tests results across all participants - fully preprocessed data | | | | | | |
|---|---|---|---|---|---|---|
| Model | Best Accuracy | Avg Accuracy | p-value | Distribution | | seed |
| | | | | Imagined | Rest | |
| Participant 2 | 0.704 | 0.677 | <0.001 | 0.70 | 0.70 | 0 |
| Participant 2 | 0.663 | 0.656 | <0.001 | 0.72 | 0.66 | 6 |
| Participant 2 | 0.745 | 0.728 | <0.001 | 0.73 | 0.71 | 12 |
| Participant 2 | 0.724 | 0.711 | <0.001 | 0.72 | 0.65 | 24 |
| Participant 2 | 0.714 | 0.697 | <0.001 | 0.74 | 0.71 | 42 |
| **Participant 2 Avg.** | **0.710** | **0.694** | **<0.001 (Fisher's test)** | **0.72** | **0.69** | **-** |
| Participant 3 | 0.694 | 0.673 | <0.001 | 0.73 | 0.66 | 0 |
| Participant 3 | 0.684 | 0.667 | <0.001 | 0.72 | 0.66 | 6 |
| Participant 3 | 0.704 | 0.677 | <0.001 | 0.70 | 0.64 | 12 |
| Participant 3 | 0.684 | 0.670 | <0.001 | 0.70 | 0.66 | 24 |
| Participant 3 | 0.704 | 0.690 | <0.001 | 0.75 | 0.68 | 42 |
| **Participant 3 Avg.** | **0.694** | **0.675** | **<0.001 (Fisher's test)** | **0.72** | **0.66** | **-** |
| Participant 4 | 0.649 | 0.622 | 0.006 | 0.69 | 0.65 | 0 |
| Participant 4 | 0.635 | 0.604 | 0.022 | 0.72 | 0.60 | 6 |
| Participant 4 | 0.662 | 0.613 | 0.007 | 0.66 | 0.60 | 12 |
| Participant 4 | 0.662 | 0.631 | 0.002 | 0.65 | 0.65 | 24 |
| Participant 4 | 0.649 | 0.626 | 0.003 | 0.63 | 0.65 | 42 |
| **Participant 4 Avg.** | **0.651** | **0.619** | **<0.001 (Fisher's test)** | **0.67** | **0.63** | **-** |
| **Tot. Avg.** | **0.685** | **0.663** | **-** | **0.70** | **0.66** | **-** |

**Supplementary Table S2.** Decoding accuracy performance of the 1D-CNN model when inputting raw data, assessed over 5 different seeds, and 3 k-folds, across all participants. Best model accuracy and average model accuracy across the different folds and different seeds per participant are reported; as well as significance ($p < 0.05$). Averaged values across participants are also outlined.

| Supplementary Table S2. 1D-CNN Model tests results with raw data across participants | | | | |
|---|---|---|---|---|
| **Participant** | **Best Accuracy** | **Avg Accuracy** | **p-value** | **seed** |
| Participant 1 | 0.381 | 0.356 | 0.56 | 0 |
| Participant 1 | 0.373 | 0.355 | 0.78 | 6 |
| Participant 1 | 0.349 | 0.312 | 0.61 | 12 |
| Participant 1 | 0.357 | 0.333 | 0.94 | 24 |
| Participant 1 | 0.408 | 0.347 | 0.32 | 42 |
| **Participant 1 Avg.** | **0.374** | **0.355** | **0.89 (Fisher's test)** | - |
| Participant 2 | 0.325 | 0.313 | 0.912 | 0 |
| Participant 2 | 0.360 | 0.330 | 0.889 | 6 |
| Participant 2 | 0.333 | 0.310 | 0.817 | 12 |
| Participant 2 | 0.307 | 0.298 | 0.609 | 24 |
| Participant 2 | 0.368 | 0.326 | 0.506 | 42 |
| **Participant 2 Avg.** | **0.339** | **0.315** | **0.977 (Fisher's test)** | - |
| Participant 3 | 0.362 | 0.300 | 0.138 | 0 |
| Participant 3 | 0.346 | 0.333 | 0.969 | 6 |
| Participant 3 | 0.331 | 0.315 | 0.892 | 12 |
| Participant 3 | 0.315 | 0.294 | 0.406 | 24 |
| Participant 3 | 0.291 | 0.273 | 0.078 | 42 |
| **Participant 3 Avg.** | **0.329** | **0.303** | **0.345 (Fisher's test)** | - |
| Participant 4 | 0.349 | 0.307 | 0.450 | 0 |
| Participant 4 | 0.349 | 0.312 | 0.555 | 6 |
| Participant 4 | 0.310 | 0.296 | 0.523 | 12 |
| Participant 4 | 0.345 | 0.307 | 0.450 | 24 |
| Participant 4 | 0.339 | 0.320 | 0.841 | 42 |
| **Participant 4 Avg.** | **0.338** | **0.308** | **0.814 (Fisher's test)** | **-** |
| **Tot. Avg.** | **0.345** | **0.320** | **-** | **-** |

**Supplementary Table S3.** Decoding accuracy performance of the 1D-CNN model when inputting minimally preprocessed (i.e., optical density) data, assessed over 5 different seeds, and 3 k-folds, across all participants. Best model accuracy and average model accuracy across the different folds and different seeds per participant are reported; as well as significance (p < 0.05). Averaged values across participants are also outlined.

| Supplementary Table S3. 1D-CNN Model tests results with optical density data across participants ||||||
| --- | --- | --- | --- | --- |
| Participant | Best Accuracy | Avg Accuracy | p-value | seed |
| Participant 1 | 0.417 | 0.412 | 0.01 | 0 |
| Participant 1 | 0.465 | 0.433 | <0.001 | 6 |
| Participant 1 | 0.433 | 0.412 | <0.001 | 12 |
| Participant 1 | 0.441 | 0.415 | <0.001 | 24 |
| Participant 1 | 0.425 | 0.497 | 0.1 | 42 |
| **Participant 1 Avg.** | **0.536** | **0.418** | **<0.001 (Fisher's test)** | - |
| Participant 2 | 0.500 | 0.444 | <0.001 | 0 |
| Participant 2 | 0.456 | 0.456 | <0.001 | 6 |
| Participant 2 | 0.509 | 0.480 | <0.001 | 12 |
| Participant 2 | 0.509 | 0.450 | <0.001 | 24 |
| Participant 2 | 0.509 | 0.439 | <0.001 | 42 |
| **Participant 2 Avg.** | **0.497** | **0.454** | **<0.001 (Fisher's test)** | - |
| Participant 3 | 0.402 | 0.368 | 0.444 | 0 |
| Participant 3 | 0.441 | 0.426 | 0.002 | 6 |
| Participant 3 | 0.409 | 0.381 | 0.189 | 12 |
| Participant 3 | 0.425 | 0.407 | 0.020 | 24 |
| Participant 3 | 0.417 | 0.398 | 0.071 | 42 |
| **Participant 3 Avg.** | **0.419** | **0.396** | **<0.001 (Fisher's test)** | - |
| Participant 4 | 0.505 | 0.484 | <0.001 | 0 |
| Participant 4 | 0.469 | 0.434 | <0.001 | 6 |
| Participant 4 | 0.452 | 0.447 | <0.001 | 12 |
| Participant 4 | 0.499 | 0.463 | <0.001 | 24 |
| Participant 4 | 0.460 | 0.458 | <0.001 | 42 |
| **Participant 4 Avg.** | **0.477** | **0.457** | **<0.001 (Fisher's test)** | **-** |
| **Tot. Avg.** | **0.482** | **0.431** | **-** | **-** |

**Supplementary Table S4.** Decoding accuracy performance of the 1D-CNN model when inputting fully preprocessed data, assessed over 5 different seeds, and 3 k-folds, across all participants. Best model accuracy and average model accuracy across the different folds and different seeds per participant are reported; as well as significance (p < 0.05). Averaged values across participants are also outlined.

| Supplementary Table S4. 1D-CNN Model tests results with fully preprocessed data across participants ||||| 
|---|---|---|---|---|
| **Participant** | **Best Accuracy** | **Avg Accuracy** | **p-value** | **seed** |
| Participant 1 | 0.405 | 0.397 | 0.057 | 0 |
| Participant 1 | 0.437 | 0.418 | 0.005 | 6 |
| Participant 1 | 0.381 | 0.368 | 0.478 | 12 |
| Participant 1 | 0.413 | 0.386 | 0.136 | 24 |
| Participant 1 | 0.445 | 0.421 | <0.001 | 42 |
| **Participant 1 Avg.** | **0.416** | **0.398** | **<0.001 (Fisher's test)** | - |
| Participant 2 | 0.518 | 0.488 | <0.001 | 0 |
| Participant 2 | 0.465 | 0.427 | 0.001 | 6 |
| Participant 2 | 0.552 | 0.497 | 0.001 | 12 |
| Participant 2 | 0.491 | 0.439 | <0.001 | 24 |
| Participant 2 | 0.491 | 0.483 | <0.001 | 42 |
| **Participant 2 Avg.** | **0.503** | **0.467** | **<0.001 (Fisher's test)** | - |
| Participant 3 | 0.441 | 0.417 | 0.006 | 0 |
| Participant 3 | 0.441 | 0.402 | 0.029 | 6 |
| Participant 3 | 0.433 | 0.387 | 0.079 | 12 |
| Participant 3 | 0.449 | 0.404 | 0.020 | 24 |
| Participant 3 | 0.417 | 0.391 | 0.085 | 42 |
| **Participant 3 Avg.** | **0.436** | **0.400** | **<0.001 (Fisher's test)** | - |
| Participant 4 | 0.492 | 0.421 | 0.001 | 0 |
| Participant 4 | 0.444 | 0.437 | <0.001 | 6 |
| Participant 4 | 0.413 | 0.397 | 0.061 | 12 |
| Participant 4 | 0.421 | 0.410 | 0.017 | 24 |
| Participant 4 | 0.452 | 0.426 | <0.001 | 42 |
| **Participant 4 Avg.** | **0.444** | **0.419** | **<0.001 (Fisher's test)** | **-** |
| **Tot. Avg.** | **0.450** | **0.421** | **-** | **-** |

**Supplementary Table S5.** Decoding accuracy performance of the Ridge Regression model when inputting raw data, assessed over 5 different seeds, and 3 k-folds, across all participants. Best model accuracy and average model accuracy across the different folds and different seeds are reported; as well as significance (p < 0.05). Averaged values across participants are also outlined.

| Supplementary Table S5. Ridge Regression Model tests results with raw data across participants | | | | |
|---|---|---|---|---|
| Participant | Best Accuracy | Average Accuracy | p-value | Seed |
| Participant 1 | 0.397 | 0.376 | 0.023 | 0 |
| Participant 1 | 0.421 | 0.360 | 0.457 | 6 |
| Participant 1 | 0.373 | 0.333 | 0.922 | 12 |
| Participant 1 | 0.357 | 0.347 | 0.505 | 24 |
| Participant 1 | 0.405 | 0.360 | 0.042 | 42 |
| **Participant 1 Avg.** | **0.390** | **0.355** | **0.075 (Fisher's test)** | - |
| Participant 2 | 0.439 | 0.427 | <0.001 | 0 |
| Participant 2 | 0.447 | 0.401 | 0.001 | 6 |
| Participant 2 | 0.395 | 0.374 | 0.028 | 12 |
| Participant 2 | 0.43 | 0.415 | <0.001 | 24 |
| Participant 2 | 0.439 | 0.409 | 0.001 | 42 |
| **Participant 2 Avg.** | **0.43** | **0.405** | **<0.001 (Fisher's test)** | - |
| Participant 3 | 0.381 | 0.357 | 0.463 | 0 |
| Participant 3 | 0.437 | 0.392 | <0.001 | 6 |
| Participant 3 | 0.341 | 0.339 | 0.712 | 12 |
| Participant 3 | 0.389 | 0.347 | 0.212 | 24 |
| Participant 3 | 0.389 | 0.360 | 0.405 | 42 |
| **Participant 3 Avg.** | **0.387** | **0.359** | **0.009 (Fisher's test)** | - |
| Participant 4 | 0.405 | 0.399 | 0.002 | 0 |

| Participant 4 | 0.444 | 0.413 | <0.001 | 6 |
| Participant 4 | 0.444 | 0.402 | <0.001 | 12 |
| Participant 4 | 0.381 | 0.373 | 0.137 | 24 |
| Participant 4 | 0.389 | 0.360 | 0.118 | 42 |
| **Participant 4 Avg.** | **0.413** | **0.389** | **<0.001 (Fisher's test)** | **-** |
| **Tot. Avg.** | **0.405** | **0.377** | **-** | **-** |

![MindPortal]**Supplementary Table S6.** Decoding accuracy performance of the Ridge Regression model when inputting minimally preprocessed (i.e., optical density) data, assessed over 5 different seeds, and 3 k-folds, across all participants. Best model accuracy and average model accuracy across the different folds and different seeds are reported; as well as significance ($p < 0.05$). Averaged values across participants are also outlined.

| Supplementary Table S6. Ridge Regression Model tests results with optical density data across participants | | | | |
|---|---|---|---|---|
| Participant | Best Accuracy | Average Accuracy | p-value | Seed |
| Participant 1 | 0.421 | 0.402 | 0.001 | 0 |
| Participant 1 | 0.444 | 0.413 | <0.001 | 6 |
| Participant 1 | 0.373 | 0.352 | 0.296 | 12 |
| Participant 1 | 0.421 | 0.402 | 0.001 | 24 |
| Participant 1 | 0.500 | 0.450 | <0.001 | 42 |
| **Participant 1 Avg.** | **0.432** | **0.404** | **<0.001 (Fisher's test)** | - |
| Participant 2 | 0.509 | 0.482 | <0.001 | 0 |
| Participant 2 | 0.491 | 0.477 | <0.001 | 6 |
| Participant 2 | 0.588 | 0.538 | <0.001 | 12 |
| Participant 2 | 0.535 | 0.509 | <0.001 | 24 |
| Participant 2 | 0.526 | 0.512 | <0.001 | 42 |
| **Participant 2 Avg.** | **0.530** | **0.504** | **<0.001 (Fisher's test)** | - |
| Participant 3 | 0.405 | 0.397 | 0.004 | 0 |
| Participant 3 | 0.389 | 0.352 | 0.580 | 6 |
| Participant 3 | 0.381 | 0.333 | 0.313 | 12 |
| Participant 3 | 0.452 | 0.423 | <0.001 | 24 |
| Participant 3 | 0.421 | 0.389 | 0.001 | 42 |
| **Participant 3 Avg.** | **0.410** | **0.379** | **<0.001 (Fisher's test)** | - |
| Participant 4 | 0.389 | 0.354 | 0.140 | 0 |
| Participant 4 | 0.405 | 0.386 | 0.005 | 6 |
| Participant 4 | 0.444 | 0.378 | <0.001 | 12 |
| Participant 4 | 0.452 | 0.410 | <0.001 | 24 |
| Participant 4 | 0.389 | 0.354 | 0.254 | 42 |
| **Participant 4 Avg.** | **0.416** | **0.377** | **<0.001 (Fisher's test)** | **-** |
| **Tot. Avg.** | **0.447** | **0.416** | **-** | **-** |

**Supplementary Table S7.** Decoding accuracy performance of the Ridge Regression model when inputting fully preprocessed data, assessed over 5 different seeds, and 3 k-folds, across all participants. Best model accuracy and average model accuracy across the different folds and different seeds are reported; as well as significance (p < 0.05). Averaged values across participants are also outlined.

| Supplementary Table S7. Ridge Regression Model tests results with fully preprocessed data across participants | | | | |
|---|---|---|---|---|
| Participant | Best Accuracy | Average Accuracy | p-value | Seed |
| Participant 1 | 0.460 | 0.399 | 0.012 | 0 |
| Participant 1 | 0.452 | 0.41 | <0.001 | 6 |
| Participant 1 | 0.413 | 0.384 | 0.006 | 12 |
| Participant 1 | 0.452 | 0.418 | <0.001 | 24 |
| Participant 1 | 0.381 | 0.376 | 0.081 | 42 |
| **Participant 1 Avg.** | **0.432** | **0.397** | **<0.001 (Fisher's test)** | - |
| Participant 2 | 0.596 | 0.564 | <0.001 | 0 |
| Participant 2 | 0.632 | 0.553 | <0.001 | 6 |
| Participant 2 | 0.667 | 0.602 | <0.001 | 12 |
| Participant 2 | 0.614 | 0.599 | <0.001 | 24 |
| Participant 2 | 0.579 | 0.553 | <0.001 | 42 |
| **Participant 2 Avg.** | **0.618** | **0.574** | **<0.001 (Fisher's test)** | - |
| Participant 3 | 0.492 | 0.455 | <0.001 | 0 |
| Participant 3 | 0.468 | 0.437 | <0.001 | 6 |
| Participant 3 | 0.452 | 0.423 | <0.001 | 12 |
| Participant 3 | 0.500 | 0.444 | <0.001 | 24 |
| Participant 3 | 0.413 | 0.405 | 0.002 | 42 |
| **Participant 3 Avg.** | **0.465** | **0.433** | **<0.001 (Fisher's test)** | - |
| Participant 4 | 0.484 | 0.397 | <0.001 | 0 |
| Participant 4 | 0.437 | 0.397 | <0.001 | 6 |
| Participant 4 | 0.429 | 0.397 | <0.001 | 12 |
| Participant 4 | 0.429 | 0.397 | <0.001 | 24 |
| Participant 4 | 0.389 | 0.378 | 0.088 | 42 |
| **Participant 4 Avg.** | **0.433** | **0.393** | **<0.001 (Fisher's test)** | **-** |
| **Tot. Avg.** | **0.487** | **0.44925** | **-** | **-** |